\documentclass[12pt]{article}
\usepackage[USenglish]{babel} %francais, polish, spanish, ...
\usepackage[T1]{fontenc}
\usepackage[ansinew]{inputenc}
\usepackage{verbatim}
\usepackage[sort,move]{cite}

\usepackage{lmodern} %Type1-font for non-english texts and characters
\usepackage{color} 

\usepackage{graphicx} %%For loading graphic files
\usepackage{amsmath}
\usepackage{amsthm}
\usepackage{amsfonts}

\usepackage{hyperref}

\usepackage{a4wide} %%Smaller margins = more text per page.
\def\a{\alpha}
\def\b{\beta}

\def\d{\delta}
\def\e{\epsilon}           % Also, \varepsilon
\def\g{\gamma}

\def\l{\lambda}
\def\m{\mu}
\def\n{\nu}
\def\o{\omega}  
\def\p{\pi}                % Also, \varpi
  \def\th{\theta}                  %     \vartheta
\def\r{\rho}                                     %     \varrho
\def\s{\sigma}                                   %     \varsigma
\def\t{\tau}

\def\D{\Delta}

\def\O{\Omega}

\def\S{\Sigma}

\def\del{\partial}              % overwritten by \nabla
\let\a=\alpha \let\b=\beta \let\g=\gamma \let\d=\delta \let\e=\epsilon
    
\let\l=\lambda \let\m=\mu \let\n=\nu  \let\r=\rho
\let\s=\sigma \let\t=\tau    
  \let\D=\Delta

\def\nn{\nonumber} \def\bd{\begin{document}} \def\ed{\end{document}}
\def\ds{\documentstyle} \let\fr=\frac \let\bl=\bigl \let\br=\bigr
\let\Br=\Bigr \let\Bl=\Bigl
\let\bm=\bibitem
\let\na=\nabla
\let\pa=\partial \let\ov=\overline
\newcommand{\be}{\begin{equation}}
\newcommand{\ee}{\end{equation}}
\def\ba{\begin{array}}
\def\ea{\end{array}}
\def\ft#1#2{{\textstyle{{\scriptstyle #1}\over {\scriptstyle #2}}}}
\def\fft#1#2{{#1 \over #2}}
\def\del{\partial}
\def\sst#1{{\scriptscriptstyle #1}}
 \def\oneone{\rlap 1\mkern4mu{\rm l}}
\def\ie{{\it i.e.\ }}
\def\via{{\it via}}
\def\semi{{\ltimes}}
\def\str{{\rm str}}
\def\tr{{\rm tr}}
\def\Dm{{{D_{\sst{max}}}}}
\def\vac{ \left | 0 \right \rangle }
\def\kvac{ \left | k \right \rangle }

\def\sp{\; \; \;}

\def\bol{ \left | B (p^+) \right \rangle}
\def\bo1{ \left | B^0 (p^+) \right \rangle}

\def\bolt{ \left | B (p^+) \right \rangle_{\t}}

\def\boxl{ \left | B (x^-) \right \rangle}
\newcommand{\bea}{\begin{eqnarray}}
\newcommand{\eea}{\end{eqnarray}}

\def\<{ \langle }
\def\>{ \rangle }

\def\S{\Sigma}

\DeclareMathOperator\arctanh{arctanh}

\renewcommand{\floatpagefraction}{0.6}
\renewcommand{\textfraction}{0.2}

\newcommand\ca{\mathcal{A}}
\newcommand\vp{\varphi}
\newcommand\beal{\begin{align}}
\newcommand\bbone{\ensuremath{\mathbbm{1}}}

\newcommand{\eq}[1]{\begin{equation}#1\end{equation}}
\newcommand{\spl}[1]{\begin{split}#1\end{split}}
\newcommand{\al}[1]{\begin{align}#1\end{align}}
\newcommand{\subeq}[1]{\begin{subequations}#1\end{subequations}}

\newcommand{\arXividhepth}[1]{\href{http://arxiv.org/abs/#1}arXiv:{\tt #1} [hep-th]}
\newcommand{\arXividother}[2]{\href{http://arxiv.org/abs/#1}arXiv:{\tt #1} [#2]}

\newcommand{\bg}[1]{\hat{#1}}
\newcommand{\wj}{\widetilde{J}}
\newcommand{\reo}{\mathrm{Re}~\!\omega}
\newcommand{\imo}{\mathrm{Im}~\!\omega}
\newcommand{\ads}{AdS_4}
\newcommand{\mcal}{\mathcal{M}}
\newcommand{\ccal}{\mathcal{C}}
\newcommand{\ncal}{\mathcal{N}}
\newcommand{\boxedeq}[1]{
\begin{equation}
\fbox{
\rule[0.7cm]{0pt}{0pt}
$#1$
\rule[-0.45cm]{0pt}{0pt}
}
\end{equation}
}

\def\d{\text{d}}

\def\slashchar#1{\setbox0=\hbox{$#1$}           % set a box for #1
\dimen0=\wd0                                 % and get its size
\setbox1=\hbox{/} \dimen1=\wd1               % get siste of /
\ifdim\dimen0>\dimen1                        % #1 is bigger
\rlap{\hbox to \dimen0{\hfil/\hfil}}      % so center / in box
#1                                        % and print #1
\else                                        % / is bigger
\rlap{\hbox to \dimen1{\hfil$#1$\hfil}}   % so center #1
/                                         % and print /
\fi}

\def\Re           {{\rm Re\hskip0.1em}}
\def\Im           {{\rm Im\hskip0.1em}}

\newcommand{\E}{\text{\tiny E}}

\begin{document}

\begin{titlepage}

\begin{center}

%\hfill ITFA-11-xx

\vskip 2cm
{\Large \bf First order formalism for the holographic duals of defect CFTs}

% \vskip 1.25 cm {\bf Geoffrey Comp\`ere$^\clubsuit$, Paul McFadden$^\spadesuit$, Kostas Skenderis$^{\spadesuit,\clubsuit,\diamondsuit}$ and Marika %Taylor$^{\spadesuit,\diamondsuit}$}
%\\ {\vskip 0.5cm \it\small
%Institute for Theoretical Physics$^\spadesuit$, Gravitation and Astro-particle Physics Amsterdam$^\diamondsuit$ and KdV Institute for Mathematics$^\clubsuit$,\\
%Science Park 904, 1090 GL Amsterdam, the Netherlands.}

\vskip 1.25 cm {\bf Yegor Korovin}
\\ {\vskip 0.5cm \it\small
KdV Institute for Mathematics,
Institute for Theoretical Physics, \\
Science Park 904, 1090 GL Amsterdam, the Netherlands.}
{\vskip 0.2cm \it \small School of Mathematical Sciences and STAG Research Centre, University of Southampton, Southampton SO17 1BJ, UK.}

{\vskip 0.2cm \small {\it E-mail: J.Korovins@uva.nl} }

\end{center}

\vskip 1 cm

\begin{abstract}
\baselineskip=16pt
We develop a first order formalism for constructing gravitational duals of conformal defects in a bottom up approach. Similarly as for the flat domain walls a single function specifies the solution completely. Using this formalism we construct several novel families of analytic solutions dual to conformal interfaces and boundaries. As a sample application we study the boundary OPE and entanglement entropy for one of the found defects.
\end{abstract}

\end{titlepage}

\setcounter{tocdepth}{2}

\tableofcontents
\pagebreak

\section{Introduction}

The conformal defects play a prominent role in many critical condensed matter systems and in the string theory giving rise to interesting physical effects \cite{Cardy:2004hm}. The majority of physical systems studied in a laboratory have defects such as boundaries or impurities. At the same time the relevant physics often exhibits the strong coupling (e.g. Kondo effect, graphene). To address such problems the holographic approach could be applied. In this paper we focus on defects of codimension one - boundaries and interfaces.

The conformal group of $d$-dimensional Minkowski space is $SO(2,d)$. A boundary CFT (BCFT) is a CFT on a space with a planar boundary which preserves the $SO(2,d-1)$ part of the conformal group. More generally, a planar interface can separate two different CFT, in which case we call it an interface CFT (ICFT). We refer to all these possibilities as defect CFTs (DCFT) and call $SO(2,d-1)$ the symmetry group of a conformal defect. Notice that in order to define a DCFT we have to start with a bulk CFT and then couple it to a defect. It turns our that the conformal symmetry on the defect still puts strong constraints on the ambient physics \cite{Cardy:1984bb, McAvity:1993ue, McAvity:1995zd}.

To realise the group of a conformal defect holographically we have to consider asymptotically $AdS_{d+1}$ space with the isometry group broken from $SO(2,d)$ down to $SO(2,d-1)$. Noting that $SO(2,d-1)$ is the isometry group of $AdS_{d}$ one is lead to consider asymptotically locally $AdS_{d+1}$ spacetimes which can be foliated with the $AdS_{d}$ slices.

Several approaches to construct holographic duals to DCFT have been pursued. References \cite{Karch:2000ct, Karch:2000gx,Erdmenger:2002ex,Skenderis:2002vf, Aharony:2003qf} used a probe $AdS_d$ brane embedded in $AdS_{d+1}$. %(see \cite{Skenderis:2002vf} for a general analysis of supersymmetric probe $AdS_d$ branes embedded in $AdS_{d+1}$). 
This brane divides the bulk into two halfs. If one imposes an orbifold condition relating the independent fluctuations in the two halves of the bulk, one ends up with a BCFT. One can also think of an $AdS_d$ brane as a spatial cutoff. The CFT on the defect in this case is conjectured to be dual to the gravity on $AdS_d$ brane. References \cite{Karch:2000ct, Karch:2000gx} were motivated by the prospects of localizing gravity on the worldvolume of a brane. The key feature allowing this localisation is the presence of the local bump of the warp factor around the position of the brane. Equivalently, the scalar potential has a characteristic vulcano shape.

%\cite{Karch:2000ct, Karch:2000gx} conjectured that a holographic dual for a boundary CFT (BCFT) is a probe $AdS_d$ brane embedded in $AdS_{d+1}$. This brane divides the bulk into two halfs. If one imposes an orbifold condition relating the independent fluctuations in the two halves of the bulk, one ends up with a BCFT. One can think of an $AdS$ brane as a spatial cutoff. 

%In our language this construction amounts to a geometry which ends at $\s_0$. 
Similar set of ideas was used to construct a holographic dual of BCFT in \cite{Takayanagi:2011zk, Fujita:2011fp}. In this approach the asymptotically locally AdS spacetime is cut off across some surface $Q$ in such a way that conformal infinity has a boundary. This additional boundary $Q$ can be realized as a brane in string theory embedding.

Regular solutions dual to ICFT describing intersecting branes and branes ending on branes have been obtained from $10$- and $11$-dimensional supergravity in \cite{D'Hoker:2007xy, D'Hoker:2007xz, Chiodaroli:2009yw, Chiodaroli:2010mv, Berdichevsky:2013ija}. These in turn in particular limits can give rise to duals of BCFTs \cite{Assel:2011xz, Aharony:2011yc, Benichou:2011aa, Chiodaroli:2011fn}. However the resulting spaces are generically singular. A non-singular example of this type was constructed in \cite{Chiodaroli:2012vc}.

%\cite{Bachas:2011xa} consider cases arising from these solutions when the central charge (radius of AdS) differs on the two sides of the interface. In type IIB context the ratio $L_-^4/L_+^4$ determines the fraction of D3-branes dissolved into the stack(s) of $5$-branes.

Finally, defect CFTs can be realized by utilising a curved domain wall ansatz for the fields where an asymptotically locally $AdS_{d+1}$ is sliced using $AdS_d$ slices. Famously, the Janus solution \cite{Bak:2003jk, Freedman:2003ax, Clark:2004sb, Papadimitriou:2004rz} is of this type. The simplicity of this solution is based on the fact that the dilatonic potential is just a constant. 

Naturally, Janus-type ansatz can be used also to construct examples involving non-trivial potential for the scalar field. In the reference \cite{Gutperle:2012hy} such solutions are constructed numerically (with unbounded potential). The BCFT solutions in \cite{Gutperle:2012hy} are singular and are conjectured to be dual to massive IR fixed point.

%If a system has more then one possible vacuum (often these are related by some symmetries) then the interfaces separating different vacua are ubiquitous.

Domain walls have been studied extensively in the context of AdS/CFT \cite{Boonstra:1998mp, Skenderis:1999mm, Bianchi:2001de, Freedman:2003ax, Papadimitriou:2004rz}. For Poincare-invariant domain walls a beautiful first order formalism  based on the so-called fake superpotential exists. Its power stems from the fact that the entire problem of finding flat domain walls reduces to solving first order ordinary differential equations (as opposed to second order partial Einstein equations). Also the stability of such domain walls is guaranteed \cite{Townsend:1984iu, Skenderis:1999mm}.

In this paper we develop further the first order formalism of \cite{Bianchi:2001de, Skenderis:2006jq, Skenderis:2006rr} for curved domain walls \footnote{An alternative first-order formalism based on introduction of two real superpotentials is given in \cite{Bazeia:2006ef}. There the superpotentials are also related by a constraint.} \footnote{An alternative solution generating technique for coupled ODEs based on single real superpotential was discussed in \cite{DeWolfe:1999cp}.}.  % \cite{Bazeia:2006ef} Gives analytic examples of (soliton-like) kink solutions connecting two vacua (both in single- and two-scalar models). Under closer inspection, geometries in these solutions are just pure AdS with some scalar profile on it (potential is tuned in such a way that scalar's back reaction vanishes).}.
In this formalism a complex superpotential (or a triplet of real superpotentials in \cite{Bianchi:2001de}) is introduced and there is a constraint relating the two independent real functions appearing in it. In this paper we point out that this constraint can be solved (at least implicitly) in a closed form. As a result, the warp factor and the scalar field satisfy first order ordinary differential equations and an $AdS$-sliced domain wall is specified completely by a single function.  In this respect the situation is now analogous to that for Poincar\'e-invariant domain walls. This observation allows us to bring the formalism in a form which can be readily applied for constructing simple examples of analytic solutions. It allows us to find families of holographic duals to boundary and interface CFTs. Moreover, all these solutions are stable \cite{Skenderis:1999mm, Skenderis:2006jq}.

The paper is organized as follows. In the next section we introduce and simplify the formalism for curved domain walls of \cite{Skenderis:2006jq, Skenderis:2006rr}. On the basis of it we construct families of analytic gravitational backgrounds in the section \ref{Examples}. As a sample application of these results we study the boundary OPE and entanglement entropy for one of the defects in the section \ref{App}.

%\cite{Galloway:1999br} claim that any Lorentzian asymptotically anti-de Sitter spacetime with a disconnected boundary-at-infinity necessarily contains black hole horizons which screen the boundary components from each other.

%If the boundary is negatively curved, slices are compact and the the p-form field strength is present then these backgrounds can suffer from non-perturbative instabilities of the type discussed in \cite{Seiberg:1999xz} (see also \cite{Maldacena:2004rf} for the discussion of this point).
%\subsection{Generalities on defect CFTs}

%Conformal group of $d$-dimensional Minkowski space is $SO(2,d)$. Boundary CFT (BCFT) is a CFT on a space with a planar boundary which preserves the $SO(2,d-1)$ part of the conformal group. More generally, a planar interface can separate two different CFT, in which case we call it interface CFT (ICFT). We refer to all these possibilities as defect CFTs (DCFT) and call $SO(2,d-1)$ the subgroup of a conformal defect. Notice that in order to define a DCFT we have to start with a bulk CFT and then couple it to a defect.  

\section{The formalism}

In this section we present the general first order formalism for domain walls with AdS slices. We begin by reviewing the construction of \cite{Bianchi:2001de, Skenderis:2006jq, Skenderis:2006rr} where a complex "superpotential" (or a triplet of real superpotentials) is introduced allowing for the first order equations for the fields. Then we demonstrate how one of the real functions appearing in the "superpotential" can be obtained from another one in the closed analytic form. This observation brings the formalism for curved domain walls in the shape similar to that for flat domain walls, where the solution is completely specified by the single superpotential. We comment on the Hamilton-Jacobi theory in Appendix \ref{HJtheory}.

\subsection{Review of the first order formalism for curved domain walls with a complex superpotential}
Let us briefly summarize the construction of \cite{Skenderis:2006jq, Skenderis:2006rr}. For simplicity we assume that the domain wall is supported by a single scalar. The theory is defined by the Lagrangian density
\be
\label{theory}
\mathcal{L} = \sqrt{-g}\Big( R - \frac{1}{2} (\pa \s)^2 - V(\s) \Big)
\ee
with general potential $V(\s)$. The formalism is applicable for both domain walls and cosmologies. It is 
convenient to introduce a sign $\eta$ such that $\eta=1$ for domain 
walls and $\eta=-1$ for cosmologies. We are looking for the solutions of the form
\be
\label{ansatz} 
ds^2_{d+1} = \eta \, dz^2 + 
e^{2\beta\varphi} ds^2_d %\left[
%- \frac{\eta\, dr^2}{1+ \eta k r^2} + r^2 d\Omega_\eta^2\right],
\ee
where $ds^2_d$ denotes the metric of a spacetime (space) of constant curvature for $\eta = 1$ ($\eta = -1$). For later convenience we introduce $d$-dependent
constants 
\be
\label{ab}
\alpha = d \beta\, , \ \qquad 
\beta =1/\sqrt{2 d(d-1)}\, . 
\ee
We use the ansatz preserving the symmetry group of the slices $ds^2_d$, i.e. the scalar field $\s$ and the warp factor $\varphi$ depend only on $z$. The field equations then reduce 
to equations for the variables $(\varphi,\s)$ that can be derived from the effective Lagrangian
\be
\label{Lagrangian}
L= \frac{\eta}{2} e^{\alpha\varphi} \left(\dot\varphi^2 -
\dot\s^2\right) 
- e^{\alpha\varphi}\left(V(\s)
- \frac{k}{2 \beta^2} e^{-2\beta\varphi}\right) \, , 
\ee
where the overdot indicates differentiation with respect to $z$ and $k \in (-1, 0, 1)$ denotes the radius of curvature of $ds_d^2$. The Euler-Lagrange equations reduce to
\be \label{feq}
\ddot\varphi = -\alpha\dot\sigma^2 -\left(k\eta/\beta\right)\, 
e^{-2\beta\varphi}\, ,\qquad
\ddot\sigma = -\alpha \dot\varphi\dot\sigma + \eta V'\, , 
\ee
where the prime indicates differentiation with respect to $\sigma$. The solutions to these equations have to satisfy an additional constraint (resulting from the Hamiltonian constraint in the original theory \eqref{theory})
\be\label{Friedman}
\dot\varphi^2 -\dot\sigma^2 =-2\eta\left[V- 
\frac{k}{2\beta^2}\, e^{-2\beta\varphi}\right]\, . 
\ee

We are looking for the solutions with the scalar field $\s$ interpolating between two critical points of the potential $V(\s)$ (these can coinside). First let us assume that such a solution $(\s(z), \varphi(z))$ is already given. If $\s(z)$ is a monotonic function we can view $\s$ as a radial coordinate by inverting the relations between $\s$ and $z$.
%If $V$ has an extremum that allows a solution with $\dot\sigma\equiv0$ 
%then the domain wall or cosmological solution is actually a
%dS, Minkowski or adS vacuum solution. So we shall 
%assume that $\dot\sigma$ is not identically zero. In fact, we shall 
%assume initially that $\dot\sigma$ is nowhere zero, returning
%subsequently to consider what happens when $\dot\sigma$ has isolated 
%zeros. Given that $\dot\sigma\ne0$, there is an inverse function 
%$z(\sigma)$ that allows any function of $z$ to be considered as a 
%function of $\sigma$. In particular,  given any solution with 
For $\eta k \le0$ we may define a complex function 
\be
Z(\sigma) = \omega(\sigma) e^{i\theta(\sigma)}
\ee
by the formulae
\bea\label{om}
\omega &=& \frac{1}{2\alpha} \sqrt{\dot\varphi^2 - 
\frac{k\eta}{\beta^2}e^{-2\beta\varphi}}\, , \\
\theta' &=&  \pm  \sqrt{-k\eta} \,  \left(\frac{\alpha}{\beta}\right)
\dot\sigma\, e^{-\beta\varphi}
\left(\dot\varphi^2 -\frac{k\eta}{\beta^2}
e^{-2\beta\varphi}\right)^{-1}.
\label{thetaprime}
\eea
Note that for flat domain wall (i.e. with $k=0$) these expressions simplify significantly, in particular $\theta(\s)$ can be set to zero.

The function $Z(\sigma)$ constructed in this way determines the scalar potential through
\be\label{VZ}
V= 2 \eta \left[|Z'|^2 -\alpha^2|Z|^2\right]\,.
\ee
Moreover the solution used to construct $Z(\sigma)$ satisfies,
\bea\label{firstorder}
&&\dot\sigma = \pm 2|Z'| \, ,
\qquad  \dot\varphi = \mp \frac{2\alpha}{ |Z'|} \,  
{\cal R}e \left(\bar Z Z'\right) \, ,\nonumber \\
&& -k\eta\, e^{-2\beta \varphi} = \left(2 \alpha \beta\, {\cal I}m\, 
\left(\bar Z Z'\right)/ |Z'|\right)^2\, . 
\eea
Importantly, these equations imply the second-order ones. We can now reverse the logic: for given $Z(\s)$ a solution of first order equations \eqref{firstorder} produces a domain wall!

The consistency between the second and third of eqs. (\ref{firstorder}) requires
\be\label{Zone}
{\cal I}m\, \left[ \bar Z' \left(Z'{}' + 
\alpha\beta Z\right)\right] =0\, . 
\ee
Remarkably,  this  is an identity  for $(\omega,\theta)$ 
defined by (\ref{om})-(\ref{thetaprime}). To summarise, any two scalar functions $\o(\s)$ and $\theta(\s)$ satisfying the constraint \eqref{Zone} define a domain wall solution (provided that the potential $V(\s)$ constructed out of them has at least one critical point).

Let us note that in the original version of this formalism \cite{Bianchi:2001de} a triplet of superpotentials $\bf{W}$ has been introduced. However as was shown in \cite{Skenderis:2006jq} the formulation with the triplet is equivalent to that with a single complex superpotential.
 
\subsection{Solving the constraint}
The consistency condition \eqref{Zone} can be solved in the closed form, i.e. given $\o(\s)$ there is an analytic formula for $\th(\s)$. \eqref{Zone} is equivalent to
\be
\label{dv}
\o \o' \th'' + ( 2 \o'^2 -\o \o''  - \a \b \o^2) \th'=-\o^2 \th'^3.
\ee
Viewed as an equation for $\th'$ this constraint can be solved in a closed analytic form. However, we first simplify it by introducing 
\be
s(\s) = {\o^2(\s)}
\ee
and
\be
\label{q}
q(\s) = \frac{1}{\o(\s)^2}(1+\frac{\o'(\s)^2}{\o(\s)^2 \th'(\s)^2}) = - k \eta \, 4 \a^2 \b^2 e^{2 \b \varphi}.
\ee
In terms of these functions equation \eqref{dv} reads
\be
\label{dq}
\frac{1}{4 \a \b} q'(\s) s'(\s) +  q(\s) s(\s) = 1.
\ee
This can be viewed as a first order linear ODE for $s(\s)$ if $q(\s)$ is given or for $q(\s)$ if $s(\s)$ is given. It has the closed form solution
\begin{align}
\label{qofs}
q(\s) &=  \exp(- 4 \a \b \int_{\s_{\star}}^{\s} \frac{s(\s')}{s'(\s')} d \s') \times \nn \\ & \qquad \qquad \times \Big( q(\s_{\star}) + 4 \a \b\int_{\s_{\star}}^{\s} \frac{1}{s'(\s')} \exp(4 \a \b \int_{\s_{\star}}^{\s'} \frac{s(\s'')}{s'(\s'')} d \s'') d\s' \Big),
\end{align}
where $\s_{\star}$ is some initial value point. Completely analogous formula holds with $s$ and $q$ interchanged.

It is amusing to note that $s(\s)$ and $q(\s)$ enter symmetrically in the constraint equation \eqref{dq}. Nevertheless their role is very different, which is evident from the expression for the scalar potential
\be
\label{V}
V(\s) = 2 \eta \Big( \frac{s'(\s)^2}{4} \frac{q(\s)}{q(\s) s(\s)-1} - \a^2 s(\s) \Big) = - 2 \eta \a \Big(\b \frac{q(\s)}{q'(\s)}s'(\s) + \a s(\s) \Big).
\ee
\eqref{q} shows that $q(\s)$ is essentially the warp factor. %We will call $s(\s)$ the superpotential. 
The scalar profile $\s(z)$ is obtained from
\be
\label{scalar}
\dot{\s}^2 = {s'(\s)^2} \frac{q(\s)}{q(\s) s(\s)-1}= - 4 \a \b \frac{q(\s)}{q'(\s)}s'(\s).
\ee
Note, that functions $s(\s)$ and $q(\s)$ must be always positive. 

Another useful equation can be derived from \eqref{firstorder}
\be
\label{qdot}
\dot{q}^2 = 16 \a^2 \b^2 q(\s(z)) \Big(q(\s(z)) s(\s(z)) -1 \Big).
\ee
It allows to obtain the warp factor as a function of the radial coordinate $z$. However we can use $\s$ to parametrize the radial direction.

\eqref{q}, \eqref{qofs}, \eqref{V} and \eqref{scalar} provide formal solution to the problem of finding AdS-sliced domain walls. I.e. given $s(\s)$ we can derive potential, geometry and the scalar profile of the solution. 

In principle one can (and would like to) view the potential $V(\s)$ as an independent function and derive an ODE for $q(\s)$ (or $s(\s)$) for any given $V(\s)$. We do not present this equation here since it is highly non-linear unilluminating second order equation which is useless in practise. 

On the practical side, the only technical difficulty in finding analytic domain-wall solutions (with prescribed properties) is hidden in evaluation of the integrals in \eqref{qofs} (or equivalently, solving first order ODE \eqref{dq} analytically).

Let us make the following comment. From the discussion above it is clear that the full solution is determined by a single function (just as for flat domain walls). We chose this free function to be $s(\s)$. But other choices are possible and can even be beneficial for constructing some analytic solutions. For example, we can parametrize solutions by the function $q(s)$. In this case $s(\s)$ as a function of scalar field can be obtained by inverting the relation
\be
\s = \pm \frac{1}{\sqrt{4 \a \b}} \int \sqrt{\frac{dq/ds}{1 - s q(s)}} ds
\ee
and the scalar profile follows from% [CHECK!]
\be
z = \pm \int \sqrt{-4 \a \b \frac{dq/ds(\s)}{q(s(\s))}} d\s.
\ee

As an example, we note that Janus solution \cite{Bak:2003jk, Clark:2004sb, Papadimitriou:2004rz} corresponds to a simple functional relation between $s$ and $q$:
\be
c \,q(s)^{-d} = s - \frac{1}{4 \a^2 \b^2 L^2},
\ee
where $L$ is the asymptotic radius of curvature and $c$ is an arbitrary constant.

Similarly, any other functional dependence between $s$, $q$, and $\s$ can be used to construct a solution generating algorithm.

\section{Examples} \label{Examples}

\subsection{Generalities}
From now on we will specialise to $AdS_{d}$-sliced domain walls in asymptotically $AdS_{d+1}$, i.e. we set $\eta = 1$, $k=-1$.

In the famous Janus solution \cite{Bak:2003jk, Freedman:2003ax, Clark:2004sb, Papadimitriou:2004rz} the potential of the scalar field $\s$ is flat producing a moduli space of vacua parametrized by the values of the dilaton. Here we are interested in the cases when potential has distinct isolated critical points. Corresponding to every critical point (maximum or minimum) of the potential there is an AdS vacuum with constant scalar $\s$ and radius of AdS determined by the value of the potential.

We are interested in finding solutions which interpolate between two asymptotically AdS region(s) corresponding to the two sides of a defect.  %corresponding to two different vacua of the potential in the case when dual field theory is an interface CFT (ICFT) or approaching the same. 
If the solution interpolates between two different vacua of the potential we interpret such solution as a bulk dual of interface CFT (ICFT). %or a dual of an RG flow (depending on the near boundary behaviour of the scalar). 
If the bulk solution is $\mathbb{Z}_2$ symmetric (i.e. we approach the same critical point on both sides of the interface) then upon identification of the two sides of AdS one obtains a holographic dual of boundary CFT (BCFT). More generally, if the bulk has only one asymptotic AdS region one can interpret it as a BCFT.

Close to the boundary of AdS (corresponding to the critical point of the scalar potential) the warp factor $q(\s)$ has to diverge. Let us make some general statements regarding the critical point. %Using \eqref{dq} we can rewrite
%\be
%\label{Vs}
%\frac{\eta}{2}V(\s) = -\a \b \frac{s'(\s) q(\s)}{q'(\s)} - \a^2 s(\s).
%\ee
Near the critical point (we can shift it to be at $\s=0$) we can expand $s(\s)$
\be
s(\s) = s_0 + \frac{s_2}{2}\s^2 + \mathcal{O}(\s^3),
\ee
the potential
\be
\label{TaylV}
V(\s) = - \frac{1}{2\b^2 L^2} + \frac{m^2}{2} \s^2 + \mathcal{O}(\s^3),
\ee
where $L$ is the radius of asymptotically AdS region and the warp factor as
\be
q(\s) = \frac{q_{-n}}{\s^n} + \frac{q_{-n+1}}{\s^{n-1}}+\mathcal{O}(\s^{-n+2}).
\ee
The constraint equation \eqref{dq} relates
\be
n = 4 \a \b \frac{ s_0}{s_2}.
\ee
There are also further relations implied by \eqref{dq}, but they depend on the particular value of $n$ and are not important for what we are going to show next. Comparing now \eqref{TaylV} with the Taylor expansion of \eqref{V} we find %the leading coefficient of the superpotential
\be
s_0 = \frac{1}{4 \a^2 \b^2 L^2}, \qquad s_2 = \frac{2(d-1)}{n L^2}.
\ee
%and 
%\be
%s_2 = \frac{2(d-1)}{n L^2}.
%\ee
%$s_2$ can take two possible values: $s_2=(d-1) \D/L^2$ and $s_2=(d-1) (d-\D)/L^2$ (where the conformal dimension of the dual operator $\D$ is determined as usually from $m^2 L^2=\D(\D-d)$). The former one correspond to the spontaneous symmetry breaking of conformal symmetry by the VEV of the dual operator, and the latter value correspond to the explicit deformation of the CFT. This discussion is analogous to that for flat domain walls [REF].
The mass of the scalar field is determined by $n$ through
\be
m^2 L^2 = -\frac{2 (d n - 2)}{n^2}
\ee
and the conformal dimension of the dual operator is determined in a remarkably simple way by $n$ as
\be
\D = \frac{1}{2}\Big(d \pm (d-\frac{4}{n}) \Big).
\ee
In particular for $n=1$ in $d=2$ the mass vanishes and we deal with dilatonic deformation (similarly as in the Janus solution \cite{Bak:2003jk, Freedman:2003ax, Clark:2004sb, Papadimitriou:2004rz}). The Breitenlohner-Freedman bound $m^2 L^2 \geq -d^2/4$ does not put any restrictions on $n$. 

Recall that for the mass in the range
\be
\label{altrange}
-\frac{d^2}{4} < m^2 L^2 < -\frac{d^2}{4} +1
\ee
there are two possible quantizations in the dual CFT which are related to each other by a Legendre transform \cite{Klebanov:1999tb}. Alternatively, one of them corresponds to Dirichlet while another to Neumann boundary condition for the scalar field on the conformal boundary of AdS. It is useful to reformulate condition \eqref{altrange} in terms of parameter $n$. For $d=2$ the alternative quantisation is possible whenever $n>1$, $n=1$ corresponds to marginal (diatonic) deformation, while $0<n<1$ leads to the non-vanishing VEV of an irrelevant operator of dimension $\D = 2/n$. For $d>2$ the alternative quantisation is possible if
\be
\frac{4}{d+2} < n < \frac{4}{d - 2}.
\ee
If $\frac{4}{d-2} < n$ then we are dealing with explicit deformation by a relevant operator of dimension $\D = d - 2/n$. % and corresponding geometries should be interpreted as RG flows (???).
For $n$ in the interval
\be
\frac{2}{d} < n <  \frac{4}{d+2}
\ee
a relevant operator of dimension $\D = 2/n$ gets a VEV, and for $0 < n <2/d$ an irrelevant operator of dimension $\D = 2/n$ gets a VEV. %Note that it seems impossible to switch on a source for an irrelevant operator and hence it is unclear if any geometry can be interpreted as an RG flow(???).

In the examples below the metric can be written explicitly using the scalar $\s$ to parametrize the radial coordinate. However, in order to understand the geometry of the solutions it is convenient to adopt the coordinate $z$ of \eqref{ansatz}. Equation \eqref{qdot} can be analyzed asymptotically as $z \rightarrow \pm \infty$, where $q$ is expected to diverge exponentially in $z$. Indeed, the leading asymptotics can be written up to an overall coefficient as %\footnote{More generally, a different prefactor can appear in the following formula. The argument given below still shows that the resulting spacetime is asymptotically AdS [MORE DETAILS?; APPENDIX?]}
\be
q(z) \sim \cosh^2 \Big(\frac{z}{L_{\pm}} \Big), \qquad \text{as} \quad z \rightarrow \pm \infty,
\ee
where the curvature radii $L_{-}$ and $L_+$ can be different for $z \rightarrow - \infty$ and $z \rightarrow  \infty$ indicating that CFTs on two sides of the interface correspond to different critical points of the potential. Asymptotically as $z \rightarrow \pm \infty$ the spacetime metric is
\be
\label{zmetric}
ds^2 \sim dz^2 + \g^2 L_{\pm}^2 \Big[\cosh^2  \Big(\frac{z}{L_{\pm}}  \Big) + \ldots \Big] \frac{du^2 - dt^2 + dr_{\parallel}^2 +r_{\parallel}^2 d \O _{d-3}^2}{u^2}, \qquad \text{as} \quad z \rightarrow \pm \infty
\ee
with some arbitrary coefficient $\g$ and $d\O_{d-3}^2$ denoting the volume element of the $(d-3)$-dimensional unit sphere. This metric is asymptotically locally AdS. It can be brought to the Fefferman-Graham form using the formulae from appendix \ref{AFG}. For $\g=1$ neglecting subleading terms in the warp factor we make a change of coordinates
\be
\label{coordchange}
 \rho = \frac{u}{\cosh  \Big(\frac{z}{L_{\pm}}   \Big)}, \qquad r_{\perp} = u \tanh  \Big(\frac{z}{L_{\pm}}  \Big),
\ee
resulting in AdS written in Fefferman-Graham (FG) coordinates
\be
ds^2 = L_{\pm}^2 \frac{d\rho^2 - dt^2 + dr_{\perp}^2 + dr_{\parallel}^2 +r_{\parallel}^2 d \O _{d-3}^2}{\rho^2}.
\ee

Importantly, the two asymptotic regions $z \rightarrow \pm \infty$ both map to conformal boundary in Fefferman-Graham coordinates, i.e. they both correspond to a theory defined in the UV. The same is true when $\g \neq 1$ (see Appendix \ref{AFG}).

Now it is natural to ask: how would an RG flow of DCFT look like in a holographic setup? To answer this question notice that in Fefferman-Graham coordinates the energy scale varies (at least asymptotically) along coordinate $\rho$. We approach UV when $\rho \rightarrow 0$ while IR corresponds to the limit $\rho \rightarrow \infty$. Borrowing intuition from flat domain walls we expect the holographic RG flow to interpolate between the near boundary ($\rho \rightarrow 0$) region of UV theory and the deep interior ($\rho \rightarrow \infty$) of the IR theory. In particular, the warp factor (in domain-wall type coordinates) should go to zero in the IR.

Comparing the FG coordinates ($\rho, x_{\perp}$) to the hyperbolic slicing coordinates ($u, z$) we see that the only way to approach the deep IR region in hyperbolic slicing coordinates is to fix $z$ and send $u \rightarrow \infty$. When we fix $z$ we restrict to a particular AdS slice. Recall that all AdS slices intersect the conformal boundary along the defect. Assuming that the dynamics on the AdS slice is dual to the dynamics on the defect theory %If we assume that each AdS slice is in some sense dual to the defect theory 
(i.e. the theory describing the subset of degrees of freedom localised on the defect) we conclude that the evolution along the FG radial coordinate $\rho$ is naturally interpreted as an RG flow of the defect theory and not as that of the ambient CFT.

We can also solve for the asymptotic behaviour of the scalar $\s$:
\be
\label{zscalar}
\s(z) \sim \frac{\s_0}{\sinh^2(\frac{z}{n L})} \sim \s_0 \Big(\frac{\rho}{r_{\perp}}\Big)^{2/n},\qquad \text{as} \quad \rho \rightarrow 0.
\ee
Note that $n$ can be different on the two sides of the defect (i.e. for $r_{\perp} >0$ and $r_{\perp}<0$).

As we have discussed above the value of $n$ determines if we are dealing with an explicit deformation or a VEV for a dual operator. Naively one would expect an RG flow if we deform the theory. To be more precise we add to the action of the CFT on each side of the interface the term with a position-dependent coupling
\be
\label{opdeform}
\int d^d x \frac{\s_0}{r_{\perp}^{d-\D}} \mathcal{O}_{\D}
\ee
which breaks the full conformal symmetry to the defect conformal group. The position-dependence of this coupling is very special. It makes this deformation marginal by power counting for any $\D$ (see \cite{Dong:2012ena} for a discussion of such couplings)! This is why an RG flow is not necessarily generated by this deformation. 

When the critical point of the potential is a local minimum the dual operator is irrelevant, i.e. $\D > d$. %The source for such an operator would diverge as $z$ approaches the boundary and therefore has to vanish. Thus we interpret the behaviour of the scalar in \eqref{zscalar} in this case as a spontaneous symmetry breaking by a position dependent VEV.
Then \eqref{zscalar} says that the symmetry is broken by a position dependent VEV of the dual operator.

In the analysis above we ignored subleading terms in \eqref{zmetric}. In general there are corrections to coordinate transformation \eqref{coordchange} which will change the position of the defect $x_{\perp}=0$. However, these corrections will not change our interpretation of the solution since the leading terms in near boundary expansions of the fields are correctly captured by \eqref{coordchange} alone (or its generalisation for $\g \neq 1$).

To summarize: at the maximum of the scalar potential the type of deformation (explicit breaking or VEV) is determined by the value of $n$ (see discussion above) while at the minimum of the scalar potential the ambient CFT is deformed by a VEV for an irrelevant operator. The interface separates two critical points of the potential deformed by position-dependent sources/VEVs. %Of course these two critical points can be connected by a Poincare invariant holographic RG flow.

Now we present some of the simplest defects which can be obtained using our formalism. 

%{\color{red}THE LAST EQUATION IS WRONG - $\s$ CAN ALSO DIVERGE AT THE BOUNDARY, I.E. SINH CAN HAVE OTHER POWER: THIS CAN LEAD TO IRRELEVANT DEFORMATION.}

%The discussion above is fine for BCFT. However, if the radius of AdS is different at two asymptotically AdS regions then this discussion has to be made independently for each side of an interface. In particular the scalar field will behave as
%\be
%\s(z) \sim \s_0 e^{\l z}.
%\ee
%The exponent $\l$ will be negative at the UV fixed point corresponding to $z \rightarrow + \infty$, while it will be positive (corresponding to irrelevant deformation) at IR fixed point corresponding to $z \rightarrow -\infty$ (see similar situation in [Bzowski et al.]). 

\subsection{Interfaces} \label{interfaces}

We begin by providing examples where $q(\s)$ has two poles. Consider
\be
\label{RGdwq}
q(\s) = \frac{e^{\s}}{\cos \s}
\ee
in $d=2$ dimensions.
Equation \eqref{dq} has general solution
\be
s(\s) = \frac{e^{-\sigma } \left(c_1+\sigma +\sin (\sigma ) \cos (\sigma )\right)}{\sin (\sigma )+\cos (\sigma )}.
\ee
Requiring $s(\s)$ to be non-singular between $-\p/2$ and $\p/2$ fixes $c_1 = (2 + \p)/4$. The potential is
\be
V(\s) =  \frac{ e^{-\sigma } (-4 \sigma \!-2 (4 \sigma \! +\pi \!+4) \sin (2 \sigma )+(4 \sigma +\pi -2) \cos (2 \sigma )+\cos (4 \sigma )-\pi -7)}{4 (\sin(\sigma )+\cos (\sigma ))^3}.
\ee
\begin{figure}[t] 
\includegraphics[width=3in, height = 2in]{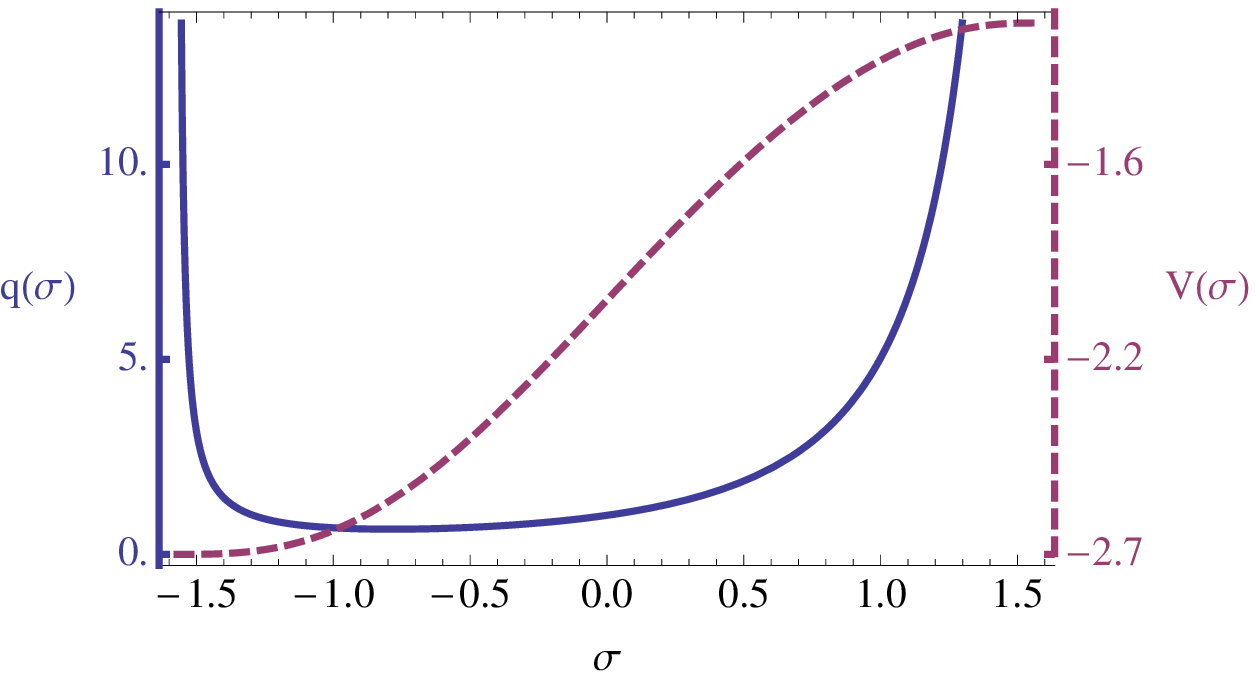}
\includegraphics[width=3in, height = 2in]{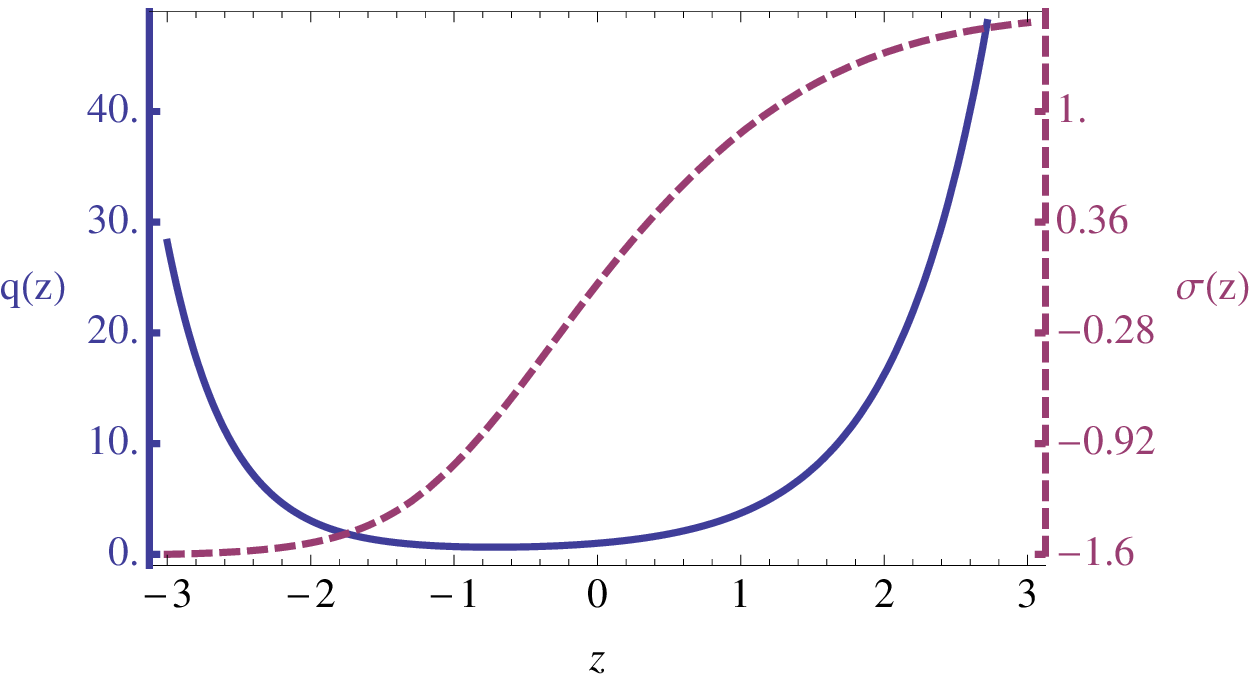}
\caption{Holographic dual for the interface CFT defined by equation \eqref{RGdwq}. Left: the warp factor and the scalar potential as functions of the scalar $\s$; Right: numerical solution for the warp factor and the scalar profile as functions of the radial coordinate $z$. Note that this interface is not $\mathbb{Z}_2$ symmetric.}
\label{RGdomwall}
\end{figure}

The potential, warp factor and the scalar profile are plotted in the Figure \ref{RGdomwall}. Between $-\p/2$ and $\p/2$ this potential is monotonic and has the maximum at $\p/2$ and the minimum at $-\p/2$.
Interestingly, the mass of the deformation at both boundaries is zero (which corresponds to $n=1$ - see the discussion above). Note that both maximum and minimum of the potential define an AdS vacuum. The radii of curvature on the two sides differ and thus the central charge in the dual theory jumps across the defect (see \cite{Bachas:2011xa} for an earlier discussion of such solution in supergravity). These two vacua could be in principle connected by an RG flow (which from the gravity perspective is just a Poincar\'e-invariant domain wall). Thus this solution is very close to a holographic realization of the construction in \cite{Gaiotto:2012np}, where such an interface is called RG domain wall. The only difference is that in our construction the theories on the two sides of the interface are deformed by position-dependent scale-invariant couplings (see previous subsection) (see also \cite{Bobev:2013yra} for a recent example of RG domain walls in supergravity).

\begin{comment}
Another similar example in $d=2$ is given by
\be
q(\s) = \frac{e^{\s}}{3-\s^2}.
\ee
Non-singular $s(\s)$ is then
\be
s(\s) = \frac{e^{-2 \sigma } \left(\frac{6992}{e}-2 e^{\sigma } (\sigma  (\sigma  (\sigma  (\sigma  ((\sigma -12) \sigma
   +63)-216)+603)-1260)+1341)\right)}{(\sigma -3)^3 (\sigma +1)}
\ee
and the potential is
\be
V(\s) = 2\eta \frac{e^{-2 \sigma -1} \left(27968 \sigma  ((\sigma -1) \sigma -3)+e^{\sigma +1} (\sigma  (\sigma  (\sigma  (\sigma  (\sigma  (\sigma  (\sigma 
   (\sigma  ((\sigma -16) \sigma +113)-504)+1782)-5256)+10422)-6120)-20727)+32184)+729)\right)}{(\sigma -3)^5 (\sigma +1)^3}.
\ee
\end{comment}

A family of interface solutions can be obtained for any $d$ from
\be
\label{interfaceq}
q(\s) = \frac{a}{\sin^n(b \s)}.
\ee
This gives %the superpotential
\be
s(\s) =\frac{1}{a}{\, _2F_1\left[-\frac{n}{2},\frac{2 \alpha  \beta }{b^2 n};\frac{2 \alpha  \beta }{b^2 n}+1;\cos ^2(b \sigma )\right]}
\ee
and the scalar potential is
\be
V(\s)=\!- \frac{2 \alpha ^2 \Big(\left(b^2 n^2 \! - \! 4 \beta ^2 \tan ^2(b \sigma )\right) \! _2F_1\!\left[\!-\frac{n}{2},\frac{2 \alpha  \beta }{b^2
   n};\frac{2 \alpha  \beta + b^2 n}{b^2 n};\cos ^2(b \sigma )\right]+4 \beta ^2 \frac{\sin ^{n+2}(b \sigma )}{\cos ^2(b \sigma )}\Big)}{a b^2
   n^2}.
\ee
For even $n$ (or simple relations among $b$ and $\b$) the hypergeometric functions reduce to polynomials (elementary functions) in $\cos ^2\sigma $. The simplest example is given by $n=2$ giving
\begin{align}
s(\s) &= \frac{1}{a} \Big(1 - \frac{\a \b}{ b^2 + \a\b} \cos^2 (b \s) \Big), \\
V(\s) &= \frac{-4 b^2 d+\cos (2 b \sigma )-1}{2 a \left(2 b^2 (d-1)+1\right)}.
\end{align}
The domain wall interpolates between $\s=0$ and $\s = \p/b$. In this example one can obtain scalar's profile and the warp factor analytically
\be
\label{Solution}
\s(z) = \frac{2}{b} \arctan \Big[ \exp (z/L) \Big], \qquad q(z) = a \cosh^2( z/L),
\ee
where we have introduced
\be
\label{Radius}
L = \frac{\sqrt{a (d - 1) (1 + 2 b^2 (d-1))}}{\sqrt{2} b}.
\ee
%\be
%\s(z) = \frac{2}{b} \arctan \Big[ \exp \Big( \frac{\sqrt{2} b z}{\sqrt{a (d-1) (1+2 b^2(d-1))}} \Big) \Big]
%\ee
%leading to
%\be
%q(z) = a \cosh^2\Big( \frac{\sqrt{2} b z}{\sqrt{a (d - 1) (1 + 2 b^2 (d-1))}} \Big).
%\ee
Note that this warp factor does not define the pure AdS space, but only asymptotically locally AdS (see Appendix \ref{AFG}).
\begin{figure}[t] 
\includegraphics[width=3in, height = 2in]{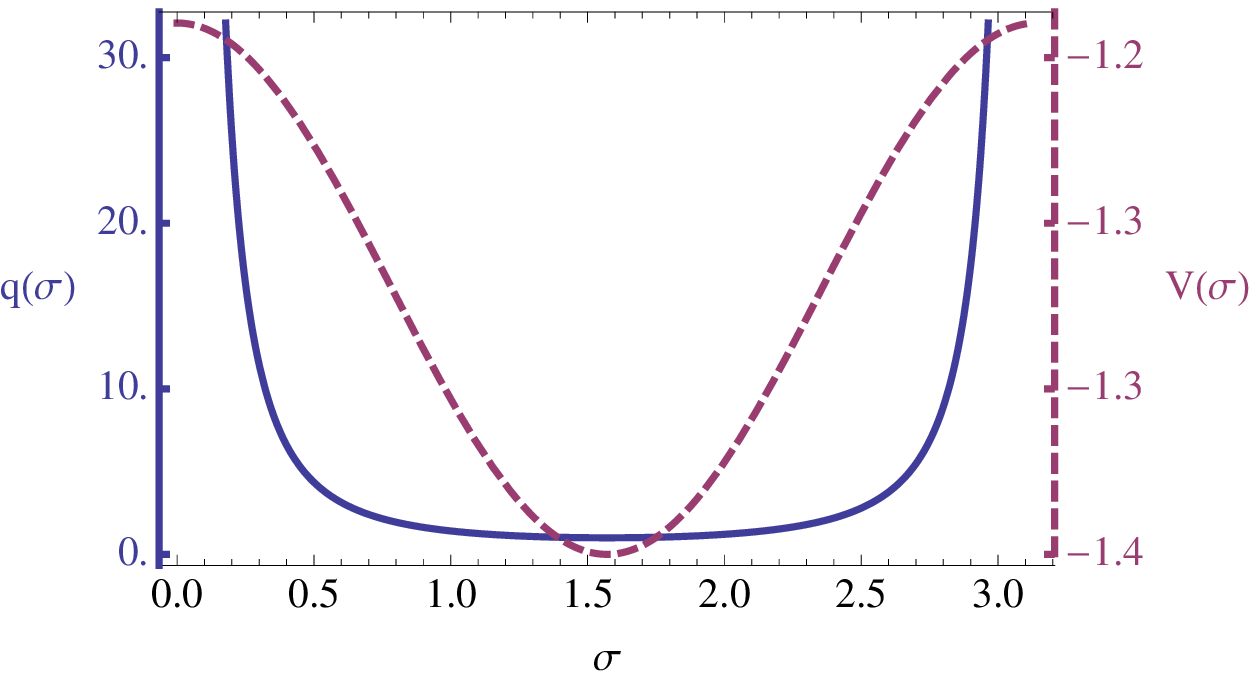}
\includegraphics[width=3in, height = 2in]{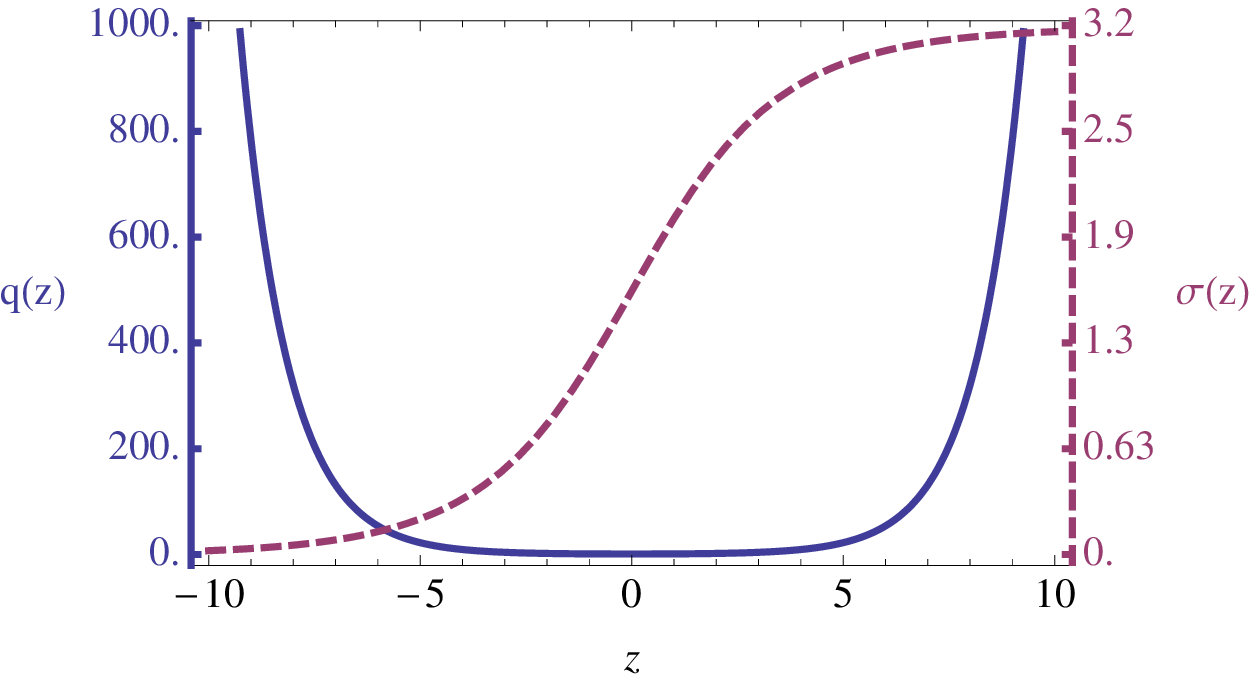}
\caption{Holographic dual for the interface CFT defined by equation \eqref{interfaceq} with $n=2$, $d=3$, $a=b=1$. Left: the warp factor and the scalar potential as functions of the scalar $\s$; Right: the warp factor and scalar profile as functions of the radial coordinate $z$.}
\label{interf}
\end{figure}

This flow interpolates between two maxima of the potential $V(\s)$ as shown on the Figure \ref{interf}.

\subsection{Boundary CFT from the folding trick}
A well-known method of constructing boundary CFT is the so-called folding (doubling) trick. With the machinery developed above it is easy to construct a holographic counterpart of this construction (see \cite{Bachas:2001vj} for an early discussion of this construction in the context of AdS/CFT correspondence). 

We want to put the same theory on the both sides of the interface. To this end it is enough if the scalar potential has only one critical point. The scalar field $\s$ approaches the same critical value as $z \rightarrow \pm \infty$. Moreover we are looking for solutions which have $\mathbb{Z}_2$ invariance with respect to $z \rightarrow-z$. It is also enough for $q(\s)$ to have only one pole.

We describe a family of solutions related to that discussed in the previous subsection. It is defined by
\be
\label{boundq}
q(\s) = \frac{a}{\sinh^n(b \s)}.
\ee
This gives %the superpotential
\be
\label{inters}
s(\s) =c_1 \cosh ^{\frac{4 \alpha  \beta }{b^2 n}}(b \sigma )+\frac{i^{-n}}{a} \, _2F_1\left[-\frac{n}{2},-\frac{2 \alpha  \beta }{b^2 n};1-\frac{2 \alpha  \beta }{b^2 n};\cosh ^2(b \sigma )\right],
\ee
where $c_1$ is an arbitrary integration constant. Note that the hypergeometric function has a branch cut in the complex plane running from $1$ to $\infty$ and generically the right hand side of the last formula should be viewed as an analytic continuation (however, for simple choices of $n$ or $b$ it reduces to elementary functions). It is easy to check then that the function $s(\s)$ defined by \eqref{inters} is in fact real. The scalar potential is
\begin{align}
&V(\s) = -\frac{8 \alpha ^2 \beta ^2 \tanh ^2(b \sigma ) \sinh ^n(b \sigma )}{a b^2 n^2} \\
&-\frac{2 \alpha ^2  \left(b^2 n^2-4 \beta ^2 \tanh ^2(b \sigma )\right) \left(i^{-n}\, _2F_1\left[-\frac{n}{2},-\frac{2 \alpha  \beta }{b^2
   n};1-\frac{2 \alpha  \beta }{b^2 n};\cosh ^2(b \sigma )\right]+a c_1 \cosh ^{\frac{4 \alpha  \beta }{b^2 n}}(b
   \sigma )\right)}{a b^2 n^2}. \nn
\end{align}

As a simple example, consider the case $n=4$ and set $c_1=0$. Clearly, as $\s$ goes to zero (from the right) we approach the boundary of AdS. The potential simplifies to
\be
V(\s) = \frac{\left(k \!-\! 1\right) \cosh (4 b \sigma )+\left(b^2 (32-48 d)+4\right) \cosh (2 b \sigma )+4 b^2 \left(d \left(11- 8 k\right)-7\right)-3}{8 a \left(k - 1\right) \left(2 k -1\right)},
\ee
where we have introduced 
\be
k = 4 b^2 (d-1).
\ee
\begin{figure}[t] 
\includegraphics[width=3in, height = 2in]{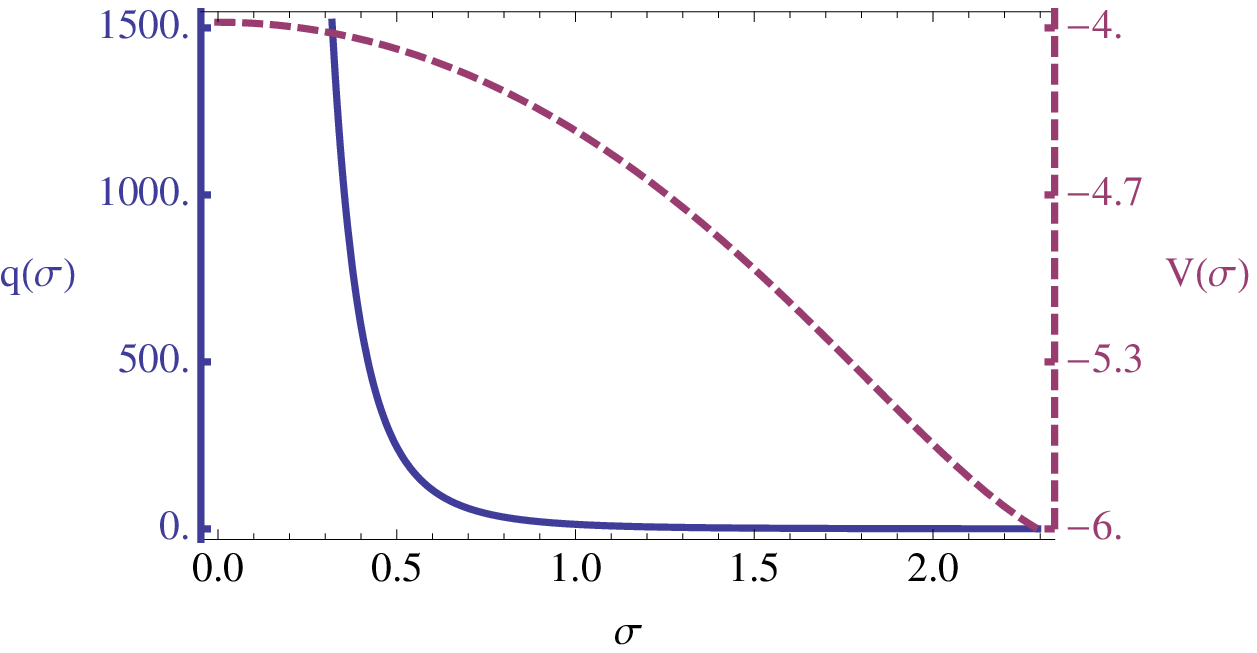}
\includegraphics[width=3in, height = 2in]{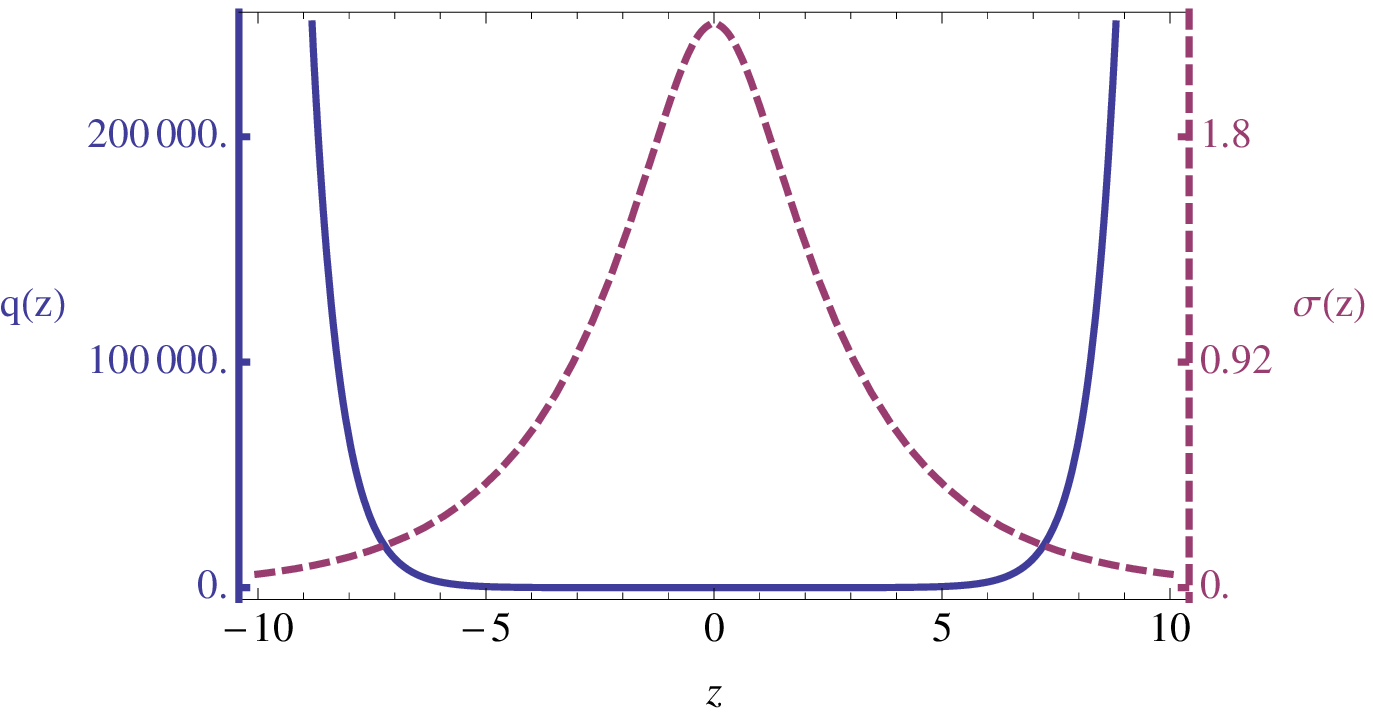}
\caption{Holographic dual for the interface CFT defined by equation \eqref{boundq} with $n=4$, $d=3$, $a=1$, $b=1/2$. Left: the warp factor and the scalar potential as functions of the scalar $\s$; Right: the warp factor and scalar profile as functions of the radial coordinate $z$.}
\label{boundary}
\end{figure}
We also impose $k>1$. The kinetic energy is
\be
\dot{\s}^2 = \frac{1}{a (d-1)}\Big(\frac{2 \sinh ^2(b \sigma )}{k-1}+\frac{\sinh ^2(2 b \sigma )}{2 \left(1-2 k\right)}\Big)
\ee
%\be
%\dot{\s}^2 =  -\frac{1}{6} \sinh ^2\left(\frac{\sigma }{2}\right) (\cosh (\sigma )-5).
%\ee
The scalar is stationary at 
\be
\s =0 \quad \text{and at} \quad \s = \frac{1}{2 b} \cosh^{-1}\Big(\frac{3 k -1}{k-1} \Big).
\ee
%\be
%\s =0 \quad \text{and at} \quad \s = \log \left(5+2 \sqrt{6}\right)
%\ee
and its profile as well as warp factor can be obtained analytically:
\be
\s(z) = \frac{1}{b} \cosh ^{-1}\Big( \sqrt{\frac{(2 k -1) }{1 + k \tanh
   ^2\left(\frac{z}{L}\right)}}\Big), \qquad q(z) = a\left( \frac{2 k - 1}{k} \cosh ^2\left(\frac{z}{L}\right) -1 \right)^2,
\ee
where the characteristic scale of this solution is (it is twice the asymptotic radius of curvature)
\be
L = \frac{\sqrt{a(k-1)(2 k -1)}}{2 \sqrt{2} b^2 }.
\ee
%\be
%\s(z) = 2 \cosh ^{-1}\left( \sqrt{\frac{3 \coth ^2\left(\frac{z}{\sqrt{6}}\right)}{\coth
%  ^2\left(\frac{z}{\sqrt{6}}\right)+2}}\right), \qquad q(z) = \left(\frac{3}{2} \cosh ^2\left(\frac{z}{\sqrt{6}}\right) -1 \right)^2.
%\ee
%The warp factor becomes
%\be
%q(z) = \left(\frac{3}{2} \cosh ^2\left(\frac{z}{\sqrt{6}}\right) -1 \right)^2.
%\ee
The potential, geometry and scalar profile of such a solution are plotted in the Figure \ref{boundary}.

Another similar family of solutions can be obtained by considering the warp factor
\be
q(\s) = \frac{a}{\s^n}.
\ee
Then
\be
s(\s) = e^{\frac{\sigma ^2}{(d-1) n}} \left(\frac{\sigma ^{n+2} \, E_{-\frac{n}{2}}\left(\frac{\sigma ^2}{(d-1)
   n}\right)}{a (d-1) n}+c_1\right),
\ee
where $E_k(x)$ denotes the exponential integral function defined as
\be
E_k(x) = \int\limits_1^{\infty} \frac{e^{-x t}}{t^k}dt.
\ee
The potential and kinetic energy can be again computed analytically analogously as in the cases discussed above. When $n$ is even the solution can be expressed using elementary functions.

\section{Applications} \label{App}

In this section we perform sample computations on some of the simplest backgrounds we have found in the previous section. We discuss the gravity dual of the boundary OPE and illustrate computation of entanglement entropy for the specific entangling surface. It allows us to extract nontrivial information about the spectrum and central charge of some defect CFTs.

\subsection{Boundary OPE and fluctuations}

In a defect CFT it is useful to draw a distinction between "ambient" fields $\phi_d(y, r_{\perp})$ propagating throughout the space and "defect" fields $\phi_{d-1}(y)$ confined to the defect ($y$ parametrizes all coordinates along the defect). The latter include also restrictions of the former to the defect.

Reference \cite{McAvity:1995zd} analysed the constraints put on the correlation functions of gauge invariant operators by the reduced defect conformal group. In particular, a non-zero one-point function for the ambient operator $\mathcal{O}_d$ of scaling dimension $\D$ is allowed and takes the form
\be
\< \mathcal{O}_d (y, r_{\perp})\> =\frac{A_{\mathcal{O}_d}}{r_{\perp}^{\D}}
\ee
for some constant $A_{\mathcal{O}_d}$. The form of the two-point function of an ambient operator of dimension $\D$ and a defect operator $\mathcal{O}_{d-1}$ of dimension $\D'$ is also fixed uniquely up to an overall coefficient
\be
\< \mathcal{O}_d (y, r_{\perp}) \mathcal{O}_{d-1} (y')\> =\frac{B_{\mathcal{O}_d \mathcal{O}_{d-1}}}{r_{\perp}^{\D - \D'} (r_{\perp}^2 + (y - y')^2)^{\D'}}.
\ee
Coordinates of any two points can be combined into an invariant with respect to the defect conformal group
\be
\xi = \frac{(y - y')^2 + (r_{\perp} - r_{\perp}')^2}{4 r_{\perp} r_{\perp}'}
\ee
and therefore the two-point function of two ambient (scalar) operators in general depends on an undetermined function $f$ of this invariant
\be
\< \mathcal{O}^1_d (y, r_{\perp}) \mathcal{O}^2_{d} (y', r_{\perp}')\> =\frac{f(\xi)}{r_{\perp}^{\D_1} r_{\perp}'^{\D_2}}.
\ee

Near the defect an ambient operator $\mathcal{O}_d$ can be expanded as a power series in defect operators $\mathcal{O}^n_{d-1}$
\be 
\label{bope}
\mathcal{O}_d (y, r_{\perp}) = \sum\limits_n \frac{B^{\mathcal{O}_d }_{\mathcal{O}^n} }{r_{\perp}^{\D_d - \D_n}} \mathcal{O}^n_{d-1}(y).
\ee
This expansion is called boundary OPE (BOPE).

As argued in the beautiful paper \cite{Aharony:2003qf} the BOPE has simple counterpart in the dual gravitational description. Consider generic field $\phi_{d+1}(z, u, y)$ of mass $M_{d+1}$ in $AdS_{d+1}$ ($y$ parametrizes coordinates along the defect while $z$ and $u$ together combine into $r_{\perp}$). It transforms in some representation of $SO(2,d)$ and is dual to some gauge-invariant ambient operator $\mathcal{O}_d$ in the boundary field theory. We can decompose $\phi_{d+1}(z, u, y)$ into a tower of fields $\phi_{d,n}(u, y)$ living on $AdS_d$ subspace and transforming in some representations of the defect subgroup $SO(2,d-1)$:
\be
\label{gravbope}
\phi_{d+1}(z, u, y) = \sum\limits_n \psi_n (z) \phi_{d,n}(u, y).
\ee
Equation \eqref{gravbope} is conjectured to be the gravitational analog of the BOPE \eqref{bope}.
The conformal dimension of $\phi_{d,n}(u, y)$ is determined by the $AdS_d$-mass $m_n^2$:
\be
\label{KG}
\Box_d \phi_{d,n}(u, y) = m_n^2 \phi_{d,n}(u, y).
\ee
After equation \eqref{KG} has been solved, the fluctuation equation for $\phi_{d+1}(z, u, y)$ reduces to an equation for $\psi_n (z)$. Imposing regularity of $\psi_n (z)$ at the boundary of $AdS_{d+1}$ (i.e. as $z \rightarrow \pm \infty$) we obtain an eigenvalue problem, which determines the masses $m_n$ of the fields appearing in the expansion \eqref{bope}. In the language of the dual field theory this analysis determines the dimensions of the operators appearing in the BOPE \eqref{bope}. 

Below we perform this program in a simple example. As a result we obtain partial information about the spectrum of the defect operators. 

First we consider the fluctuations of the minimally coupled probe scalar field on the generic backgrounds constructed above. Recall that the fluctuations of the transverse traceless part of the metric satisfy the field equation of the massless scalar (see \cite{Bachas:2011xa} for a recent discussion). We will be able to restrict to this particular case later by setting $AdS_{d+1}$-mass to zero.

For the scalar field $f(\s,u,t,\vec{x})$ of mass $M_{d+1}$ the field equation is
\begin{align}
-\frac{2 \a \b}{q'^2} \Big[ 2 q q' s' f'' &+ \Big( (d+1)s' q'^2 - (q q'' s' - q q' s'') \Big) f' \Big]  \\
&+ \frac{4 \a^2 \b^2}{q} \Big[u^2 \pa_u^2 f + (2-d)  u \pa_u f - u^2 k^2 f \Big] = M_{d+1}^2 f, \nn
\end{align}
where $k$ is the Fourier momentum along the $(t,\vec{x})$ directions. The second line involves the Laplacian in transverse $AdS_d$ space. We can separate the variables by decomposing the general fluctuation in the sum over those respecting the defect conformal group
\be
\label{decomposition}
f(\s,u) =\sum_n \psi_n(\s) U_n(u, t , \vec{x}).
\ee 
Functions $U_n$ satisfy the field equation of the probe scalar on $AdS_d$ space:
\be
u^2 \pa_u^2 U + (2-d)  u \pa_u U - u^2 k^2 U = m_n^2 U,
\ee
with the regular solution given in terms of modified Bessel function
\be
U_n (u) = u^{\frac{d-1}{2}} K_{\n_n} (k u),
\ee
where $\n_n = \sqrt{(d-1)^2 + 4 m_n^2}/2$. The field equation for the wave function $\psi_n(\s)$ is
\be
\label{wfeq}
-\frac{2 \a \b}{q'^2} \Big[ 2 q q' s' \psi_n'' + \Big( (d+1)s' q'^2 - (q q'' s' - q q' s'') \Big) \psi_n' \Big] = (M_{d+1}^2 - \frac{4 \a^2 \b^2}{q}m_n^2) \psi_n.
\ee
Generically this equation does not have an analytic solution. We can look for specific choices of $s(\s)$ for which this equation can be solved.

As a simple example consider
\be
q = \frac{a}{\sin^2(b \s)}
\ee
discussed in the section \ref{interfaces}.
%leading to periodic potential
%\begin{align}
%s(\s) &= \frac{1}{a} \Big(1 - \frac{\a \b}{ b^2 + \a\b} \cos^2 (b \s) \Big), \\
%V(\s) &= \frac{-4 b^2 d+\cos (2 b \sigma )-1}{2 a \left(2 b^2 (d-1)+1\right)}.
%\end{align}
%The domain wall interpolates between $\s=0$ and $\s = \p/b$. In this example one can obtain scalar's profile analytically
%\be
%\s(z) = \frac{2}{b} \arctan \Big[ \exp \Big( \frac{\sqrt{2} b z}{\sqrt{a (d-1) (1+2 b^2(d-1))}} \Big) \Big]
%\ee
%leading to
%\be
%q(z) = a \cosh^2\Big( \frac{\sqrt{2} b z}{\sqrt{a (d - 1) (1 + 2 b^2 (d-1))}} \Big).
%\ee
%Note that this warp factor does not define the pure AdS space, but only asymptotically AdS.
To analyse the equation \eqref{wfeq} we introduce new variable $x = \cos (b \s)$ and look for the solution of the form $\psi_n = (1-x^2)^{d/4} g_n(x)$. The equation for $g_n(x)$ becomes that for the associated Legendre function
\be
(1-x^2) g_n''(x) - 2 x g_n'(x) + \Big[\l_n (\l_n+1) -\frac{\m^2}{1-x^2} \Big]g_n(x) = 0,
\ee
with the prime denoting the derivative with respect to $x$. We have also introduced parameters $\l_n$ and $\m$ defined by
\begin{align}
(\l_n+\frac{1}{2})^2 &= \frac{(d-1)^2}{4} +  (1 +\frac{\a \b}{b^2}) m_n^2,\\ %= \frac{(d-1)^2}{4} + l^2 m_n^2 = \frac{(d-1)^2}{4} + \D_n (\D_n - (d-1))
\m^2 &= \frac{d^2}{4} + (1 +\frac{\a \b}{b^2}) \frac{a}{4 \a^2 \b^2} M_{d+1}^2.% = \frac{d^2}{4} + L^2 M_{d+1}^2 = \frac{d^2}{4} + \D (\D-d).
\end{align}
%where we have introduced the effective radius of curvature $l$ along the slice. 
We are looking for a solution for $\psi_n$ which approaches zero at the boundary, i.e. as $x \rightarrow \pm 1$. Using the trigonometric expansions for the Legendre functions \cite{bateman1953higher} we see that this condition leads to the relation
\be
\l_{n} = - \m + n, \qquad n \in \mathbb{Z}.
\ee
If $\m$ is half-integer we have to impose in addition $\l_n > -3/2$. The dimensions of the operators appearing in the BOPE decomposition are
\be
\D_n = \frac{d + \sqrt{d^2 + \frac{b^2}{b^2 + \a \b} (2 \l_n + d)(2 \l_n + 2 -d)}}{2}.
\ee
Putting $M_{d+1} = 0$ (leading to $\m$ being integer or half-integer) we obtain part of the spectrum of the operators appearing in the BOPE of the stress-energy tensor with the defect. As a simple check of our result we see that in the limit when $b \rightarrow \infty$ (which corresponds to pure AdS without defect) we get the mass spectrum
\be
m_n^2 = \frac{(2 n + d)(2 n + 2 - d)}{4}
\ee
reproducing the known result (see reference \cite{Karch:2000ct} for pure $AdS_5$ case).

\subsection{Entanglement entropy in CFTs with defects}

Entanglement Entropy (EE) in QFT is a powerful measure of entanglement between subsystems. It can count number of degrees of freedom, characterise phases of matter, serve as an order parameter in phase transitions, etc. However, EE is difficult to compute, even in free field theories. In the context of holography a well-defined prescription was proposed in \cite{Ryu:2006bv, Ryu:2006ef}. Rigorous proof of this conjecture exists in $d=2$ \cite{Hartman:2013mia, Faulkner:2013yia} (for more general evidence see \cite{Lewkowycz:2013nqa}). In the special case when the entangling surface (i.e. surface separating a subsystem from its complement) is a sphere (or a conformal transformation of it) an independent derivation of EE was given by Casini, Huerta and Myers (CHM) in \cite{Casini:2011kv}. The CHM idea was generalised for defect CFTs to compute the EE across the spherical entangling surface centred on the defect in \cite{Jensen:2013lxa}. In this sections we compute the EE of such a region in one of our examples.

Let us start with some general observations. In our notation the backreacted geometries take the form
\be
ds^2 = dz^2 + (d-1)^2  \frac{q(z)}{u^2} (du^2 - dt^2 + dr_{\parallel}^2 + r_{\parallel}^2 d\O_{d-3}^2).
\ee

We are looking for the minimal surface ending on the boundary on the circle of radius $R$ centred at the defect. We parametrize it by the so far unknown function $u(z,  r_{\parallel})$.
The minimal area functional is (assuming $d \geq 3$)
\be
\mathcal{A}_{\text{min}} = \text{Vol}(S^{d-3}) \int dr_{\parallel} dz \; r_{\parallel}^{d-3} \Big(\frac{(d-1)^2 q(z)}{u^2} \Big)^{\frac{d-2}{2}} \sqrt{1 \!+\! (\pa_{r_{\parallel}}u)^2 \!+\! \frac{(d-1)^2 q(z)}{u^2} (\pa_z u)^2}.
\ee
Crucially, as argued in detail in \cite{Jensen:2013lxa}, the area is minimised (globally) when $\pa_z u =0$. This simplifies the computation significantly. The minimal surface is given by the equation
\be
u(r_{\parallel})^2 + r_{\parallel}^2 = R^2
\ee
and its area factorizes into two integrals
\be
\label{minsurface}
\mathcal{A}_{\text{min}} = (d-1)^{d - 2} \text{Vol}(S^{d - 3}) \Big [\int\limits_0^{r_{\e}} dr_{\parallel} \frac{R \, r_{\parallel}^{d-3}}{(R^2 - r_{\parallel}^2)^{\frac{d-1}{2}}}\Big] \Big[ \int\limits^{z_{\e}(r_{\parallel})}_{-z_{\e}(r_{\parallel})} dz q(z)^{\frac{d-2}{2}} \Big],
\ee
i.e. for any fixed value of $r_{\parallel}$ between zero and $r_{\e} = \sqrt{R^2 - \e^2}$ we integrate over $z$ up to the cutoff $z_{\e}(r_{\parallel})$ which depends on both $\e$ and $r_{\parallel}$.
%The cutoff $z_{\e}$ depends on $r_{\parallel}$, i.e. 

Usually entanglement entropy is computed with a regulator $\e$ in Poincar\'e coordinates. It is possible to implement it directly in AdS slicing using the expression for the Poincar\'e radial coordinate $\r$ \cite{Jensen:2013lxa} (see also Appendix \ref{AFG})
\be
\rho = u \exp \Big(-\int^z dz' \sqrt{\frac{1}{{ L^2}}-\frac{1}{(d-1)^2 q(z)}} \Big)
\ee
leading to the relation between the cutoffs $\e$ and $z_{\e}$ %in different coordinates systems
\be
\label{cutoff}
\e = \sqrt{R^2 - r_{\parallel}^2} \exp \Big(-\int^{z_{\e}} dz' \sqrt{\frac{1}{{ L^2}}-\frac{1}{(d-1)^2 q(z)}} \Big).
\ee
After the integral over $z$ in \eqref{minsurface} is performed one integrates over $r_{\parallel}$ from $0$ to $r_{\e} = \sqrt{R^2 - \e^2}$.

We perform this computation for the interface discussed in the previous section defined by the equations \eqref{Solution} and \eqref{Radius}.
%\be
%q(\s) = \frac{a}{\sin^2(b \s)} \quad \rightarrow \quad q(z) = a \cosh^2(z/L),
%\ee
%where we have introduced
%\be
%L = \frac{\sqrt{a (d - 1) (1 + 2 b^2 (d-1))}}{\sqrt{2} b}.
%\ee
To evaluate the $z$-integral in \eqref{minsurface} we notice that
\[ \int \cosh^{d-2}(z) dz = \left\{ 
  \begin{array}{l l}
    \frac{1}{2^{d-2}} {d-2 \choose (d-2)/2} z + \frac{1}{2^{d-2}} \sum\limits_{k=0}^{\frac{d}{2}-2} {d-2 \choose k} \frac{\sinh((d-2-2k)z)}{\frac{d}{2}-1-k} & \quad \text{if $d$ is even}\\
    \sum\limits_{k=0}^{\frac{d-3}{2}} {\frac{d-3}{2} \choose k} \frac{\sinh^{2k+1}(z)}{k+1} & \quad \text{if $d$ is odd}
  \end{array} \right.\]

For definiteness we choose $d=3$, then
\be
\label{Minsurface}
\mathcal{A}_{\text{min}} = 4 R L \sqrt{a} \int \limits_0^{r_{\e}} \frac{\sinh(z_{\e}(r_{\parallel})/L)}{R^2 - r_{\parallel}^2} d r_{\parallel}.
%\int^{z_{\e}}_{-z_{\e}} dz \, q(z)^{\frac{1}{2}} = 2 L \sqrt{a} \sinh(z_{\e}/L).
\ee
Next step is to find the relation between $\e$ and $z_{\e}$. Evaluating the integral in \eqref{cutoff} we find
\be
\label{cutrelation}
\e = \sqrt{R^2 - r_{\parallel}^2} \Big( \frac{t_{\e}-1}{t_{\e}+1} \Big)^{1/2} \Big( \frac{A \, t_{\e} + 1}{A \, t_{\e} - 1} \Big)^{A/2}, 
\ee
where
\be
A^2 = \frac{L^2}{a (d-1)^2}
\ee
and we have introduced 
\be
t_{\e} = \frac{\sinh(z_{\e}/L)}{\sqrt{\cosh^2(z_{\e}/L) - A^2}}.
\ee
For generic value of $A$ one cannot solve analytically the algebraic equation \eqref{cutrelation} for $\sinh(z_{\e}/L)$ (except for the case $A=2$). Luckily, we know the inverse relation between $r_{\parallel}$ and $t_{\e}$ and can express the integrand in \eqref{Minsurface} in terms of $t_{\e}$:
\be
\label{surface}
\mathcal{A}_{\text{min}} = 4 R L \sqrt{a} (A^2-1)^{3/2} \int\limits_{t_0}^{\infty} \frac{t_{\e}}{(t_{\e}^2 - 1)^{3/2}} \frac{1}{A^2 t_{\e}^2 -1} \frac{1}{\sqrt{R^2 - \e^2 \frac{t_{\e} + 1}{t_{\e} -1} (\frac{A t_{\e} -1}{ A t_{\e} + 1})^A}} d t_{\e},
\ee
where $t_0$ is defined through the equation
\be
\frac{t_0 - 1}{ t_0 + 1} \Big(\frac{A t_0 + 1}{A t_0 -1}\Big)^A = \frac{\e^2}{R^2},
\ee
which again cannot be solved analytically for generic $A$. It is enough however to know the Taylor series for $t_0(\e)$:
\be
t_0 \sim 1 + \frac{2}{R^2} \Big(\frac{A-1}{A+1}\Big)^A \e^2+ \mathcal{O}(\e^4).
\ee

The integral in \eqref{surface} cannot be performed analytically, but we can analyse its behaviour as $\e \rightarrow 0$. Naively we expect to find $1/\e$ divergence characteristic for three-dimensional CFT and potential logarithmic divergence coming from the two-dimensional defect. 

It is convenient to change the integration variable once again and define
\be
t = t_{\e} - t_0,
\ee
then the integration domain for $t$ is from zero to infinity.
Next, let us rewrite the integral in \eqref{surface} as a sum of two integrals: one from zero to some intermediate value $\delta$ and the second one from $\delta$ to infinity (we will assume that $\delta$ is of order $\mathcal{O}(1)$ and that $\e^2 \ll \delta$). Now note that the integrand is approaching zero as $t \rightarrow \infty$ fast enough and therefore the second integral is never divergent irrespectively of the value of $\e$, even for $\e=0$. Therefore we can concentrate on the region close to the lower limit of the integral.

Notice also that the expression under the square root in the integrand vanishes at $t=0$ (as $t/\e^2$) and is of order $\mathcal{O}(1)$ when $t \sim \mathcal{O}(\e^2)$. Thus we can identify two different regions between zero and $\delta$ where the integrand behaves differently. For $t < \e^2$ it goes like $1/\sqrt{t}$ and for $\e^2 < t <\delta$ it is approximately $1/t^{3/2}$. In both regimes the power is fractional and therefore there can be no logarithmic divergence. To be more precise, in the first regime we get
\be
\int\limits_0^{\e^2} \frac{1}{\e^2} \frac{1}{\sqrt{t}} \sim \frac{1}{\e},
\ee 
and in the second regime
\be
\int\limits^{\delta}_{\e^2} \frac{1}{(t + t_0-1)^{3/2}} \sim \frac{1}{\sqrt{t + t_0-1}} \Big|_{\e^2} \sim \frac{1}{\e}.
\ee
The coefficient of $1/\e$ divergence (as well as finite part of the EE) can be extracted numerically.

Thus we do not find a logarithmic divergence. Usually the logarithmic divergence in the EE in even dimensions is multiplied by some linear combination of the central charges. For instance in a two-dimensional CFT there is a single central charge $c$ which can be defined as a coefficient of the Ricci scalar $R$ in the trace anomaly. When we are dealing with a two-dimensional defect embedded into some three-dimensional CFT the situation is more complicated. Apart from Ricci scalar one could construct other scalars which might appear in the trace anomaly, e.g. by considering different contractions of extrinsic curvatures or normal components of the metric \cite{Jensen:2013lxa}. The number of possible central charges characterising two-dimensional defects is not known. In general some linear combination of them will multiply the logarithmic divergence in the EE. Our computation shows that for this specific interface some combination of central charges vanishes. %Similar result was obtained for the Janus interface in ABJM \cite{Andrew}.

\section{Conclusions}
To summarise, we have developed a simple first-order formalism for constructing $AdS_{d}$-sliced domain walls in asymptotically $AdS_{d+1}$ spacetimes. Such domain walls provide holographic duals of the conformal codimension one defects - boundaries and interfaces. We have shown that the entire solution is specified with the help of a single function, from which the spacetime geometry, scalar potential and scalar profile are easily derived. In this respect the situation is analogous to the well-known fake superpotential formalism for flat Poincar{\'e}-invariant domain walls.

We have applied this formalism to construct families of analytic gravitational duals of the conformal defects. Among them there are subfamilies given in terms of elementary functions. For one of the simplest solutions we have computed the discrete spectrum of dimensions appearing in the OPE between the energy-momentum tensor and the defect. For the same defect we have studied the entanglement entropy across the spherical region centred on the defect. For a two-dimensional defect we have found that the coefficient in front of the logarithmic divergence in the entanglement entropy vanishes. This means that one of the  
linear combinations of the central charges vanishes in the dual defect theory.

%studied boundary OPE and entanglement entropy across the spherical region centred on the defect. We have computed the discrete spectrum of dimensions appearing in the OPE between the energy-momentum tensor and the defect. For a two-dimensional defect we have found that the coefficient in front of the logarithmic divergence in the entanglement entropy vanishes. This means that one of the  
%linear combinations of the central charges vanishes in the dual theory.% [EXPLAIN BETTER].

Interfaces constructed using the $AdS$-slicing have the property that some of the sources or VEVs have specific position-dependence of the form \eqref{opdeform}, which in part may be responsible for the breaking of bulk conformal symmetry to the defect conformal group. It is important to find a generalization of the original ansatz such that position-independent couplings could be incorporated. In such a setup it will be natural to look for a holographic version of the $g$-theorem \cite{Affleck:1991tk}. 

We have also found a simple interface separating two CFTs which could be connected via an RG flow. Such an interface is reminiscent of the construction in \cite{Gaiotto:2012np} {\footnote {We are aware that Davide Gaiotto has unpublished related work \cite{Gaiotto}.}}. The main difference in our solution is that the UV and IR CFTs on the two sides of the interface are deformed by position-dependent (but classically scale invariant) couplings or VEVs. It would be interesting to find a holographic RG flow interpolating between a CFT in the UV and such a defect in the IR.

Using domain wall / cosmology correspondence \cite{Skenderis:2006jq, Skenderis:2007sm} our formalism can also be applied for constructing cosmologies with spherical spatial slices and time-dependent cosmological constant.

\section*{Acknowledgments}

I am grateful to Kostas Skenderis for suggesting this project, numerous helpful discussions and comments on the draft. I also would like to thank Marco Meineri and Ioannis Papadimitriou for useful discussions and Andrew O'Bannon for the comments on the draft. I acknowledge support via an NWO Vici grant of Kostas Skenderis.  

\appendix

\section{Relation to the Hamilton-Jacobi Theory } \label{HJtheory}

The Hamilton-Jacobi theory for (curved) domain walls was developed in the reference \cite{Skenderis:2006rr} (see also \cite{Papadimitriou:2007sj, Papadimitriou:2010as}). Here we repeat their analysis in our language. 

Starting from effective Lagrangian \eqref{Lagrangian} we define canonical momenta
\be
\p_{\varphi} = \frac{\pa L}{\pa \dot{\varphi}} = \eta e^{\a \varphi} \dot{\varphi}, \qquad \p_{\s} = \frac{\pa L}{\pa \dot{\s}} = -\eta e^{\a \varphi}  \dot{\s}.
\ee

The Hamilton's characteristic function $\mathcal{S}$ is a functional of the fields $\s$ and $\varphi$. We will find it without actually solving the Hamilton-Jacobi equations. To this end observe that 
\be
\dot{\mathcal{S}} = \p_{\varphi} \dot{\varphi} + \p_{\s} \dot{\s}.
\ee
Using explicit expressions
\be
\dot{\varphi} = \mp 2 \a \sqrt{s - 1/q}, \qquad \dot{\s} = \pm |s'| \sqrt{\frac{1}{s - 1/q}}
\ee
we compute
\begin{align}
\eta \dot{\mathcal{S}} &=  e^{\a \varphi} (\dot{\varphi}^2 - \dot{\s}^2) \\
&= \dot{\varphi} (\mp 2 \a e^{\a \varphi} \sqrt{s - 1/q}) - \dot{\s}^2 e^{\a \varphi} \nn \\
&= \frac{d}{d z} (\mp 2 e^{\a \varphi} \sqrt{s - 1/q}) \pm \frac{e^{\a \varphi}}{\sqrt{s - 1/q}} \Big( \dot{s} +\frac{\dot{q}}{q^2}\Big) - \dot{\s}^2 e^{\a \varphi} \nn \\
& = \frac{d}{d z} (\mp 2 e^{\a \varphi} \sqrt{s - 1/q}) \mp 4 \a \b \frac{e^{\a \varphi}}{q}. \nn
\end{align}
Integrating with respect to $z$ we finally find
\be
\label{charfn}
\mathcal{S} = \mp 2  \eta e^{\a \varphi} \sqrt{s - 1/q} \mp 4  \eta  \a \b \int^{z(\s)} \frac{e^{\a \varphi}}{q} dz.
\ee
It is straightforward to check that the relations
\be
\frac{\pa \mathcal{S}}{\pa \varphi} = \p_{\varphi}, \qquad \frac{\pa \mathcal{S}}{\pa \s} = \p_{\s}
\ee
hold.

As was noted in \cite{Skenderis:2006rr} the last term in \eqref{charfn} is absent for flat domain walls. Importantly, this last term is a non-local function of the scalar field $\s$ ($\varphi$ dependence can be removed by a gauge choice). In the context of holography $\mathcal{S}$ plays the role of the generating functional and thus may need to be renormalized. While the first term can be removed by a local counterterm, the second term can not. Therefore the $1$-point function of the operator dual to $\s$ is generically non-zero. 

\section{From AdS slicing to Fefferman-Graham} \label{AFG}

A locally asymptotically $AdS_{d+1}$ metric in $AdS_d$ slicing
\be
\label{metric}
ds^2 = dz^2 + (d-1)^2 \frac{q(z)}{u^2} \Big( du^2 - dt^2 + dr_{\parallel}^2 + r_{\parallel}^2 d\O_{d-3}^2 \Big)
\ee
can be always brought in the Fefferman-Graham form (at least near the boundary and away from the location of the defect)
\be
\label{FG}
ds^2 = \frac{L^2}{\r^2} \Big[ d\r^2 + v(\r/r_{\perp}) dr_{\perp}^2 + w(\r/r_{\perp})\Big(-dt^2 + dr_{\parallel}^2 + r_{\parallel}^2 d\O_{d-3}^2 \Big)  \Big]
\ee
by the following change of coordinates \cite{Estes:2012nx, Jensen:2013lxa}
\bea
\r &=& u \, \exp\Big(\mp \int^z \sqrt{\frac{1}{L^2} - \frac{1}{(d-1)^2 q(z')}} dz'\Big), \\
r_{\perp} &=& u \, \exp\Big(\pm \frac{1}{{(d-1)}}\int^z \frac{1}{\sqrt{q(z') \Big(\frac{(d-1)^2}{L^2}q(z') -1 \Big)}}dz'\Big),
\eea
%where the sign in the first formula should be chosen in such a way that $\r$ approaches zero as $z \rightarrow \pm \infty$. Moreover
%\bea
%v(z) &=& \frac{\r^2}{r_{\perp}^2} \Big(\frac{(d-1)^2}{L^2}q(z) - 1 \Big), \\
%w(z) &=& \frac{(d-1)^2}{L^2} \frac{\r^2}{u^2} q(z),
%\eea
%where we view $z$ as a function of $\r/r_{\perp}$. We also impose
%\be
%\lim_{\r \rightarrow 0} w(\r/r_{\perp}) = 1
%\ee
%and analogous condition for $v$. This actually fixes value of the asymptotic radius of curvature $L$ in \eqref{FG} (which could have been deduced directly from \eqref{metric} of course).
where the sign in the first formula should be chosen in such a way that $\r$ approaches zero as $z \rightarrow \pm \infty$ (the sign in the second formula is then the opposite one). Moreover the integration constant is fixed by the requirement that the functions
\bea
v(z) &=& \frac{\r^2}{r_{\perp}^2} \Big(\frac{(d-1)^2}{L^2}q(z) - 1 \Big), \\
w(z) &=& \frac{(d-1)^2}{L^2} \frac{\r^2}{u^2} q(z),
\eea
approach one as $\r \rightarrow 0$.

%where we view $z$ as a function of $\r/r_{\perp}$. We also impose
%\be
%\lim_{\r \rightarrow 0} w(\r/r_{\perp}) = 1
%\ee
%and analogous condition for $v$. This actually fixes value of the asymptotic radius of curvature $L$ in \eqref{FG} (which could have been deduced directly from \eqref{metric} of course).

As a concrete example consider a geometry defined by
\be
\label{metrexample}
ds^2 = dz^2 + \g^2 L^2 \frac{\cosh^2(z/L)}{u^2} \Big( du^2 - dt^2 + dr_{\parallel}^2 + r_{\parallel}^2 d\O_{d-3}^2 \Big).
\ee
We have encountered a geometry of this type in the subsection \ref{interfaces}. When the constant $\g$ is not equal to one this space is not pure AdS. To see it compute the Ricci curvature:
\be
R = -\frac{d(d+1)}{L^2} + \frac{d(d-1) (\g^2-1)}{\g^2 L^2} \frac{1}{\cosh^2(z/L)}.
%\frac{6 (\gamma -1) \text{sech}^2\left(z/L\right)-12 \gamma }{\gamma  L^2}.
\ee
As $z \rightarrow \infty$ the Ricci scalar approaches $-d(d-1)/L^2$ which is the Ricci curvature of the pure $AdS_{d+1}$ with the radius $L$. Therefore this geometry defines asymptotically locally AdS space. Assuming that $\g^2 < 1$ the Fefferman-Graham form is achieved through
\be
r_{\perp} = u \Big(\frac{1 - \g}{1 + \g}\Big)^{\frac{1}{2 \g}}\,\Big(\frac{t+\g}{t-\g}\Big)^{\frac{1}{2 \g}},\qquad
\r = \frac{2}{\sqrt{1-\g^2}} \, r_{\perp} \Big(\frac{t - 1}{t + 1}\Big)^{\frac{1}{2}},
\ee
where
\be
t = \frac{\sinh(z/L)}{\sqrt{\cosh^2(z/L) - 1/\g^2}}.
\ee
Note that this change of coordinates is not defined on the entire spacetime, but only when $\cosh(z/L) > 1/\g$. It indicates the breaking of Fefferman-Graham expansion at a particular value of $z$.

%[TO FINISH]Finally we would like to emphasise the following technical detail. Consider a probe minimally coupled scalar field $\phi$ with mass $M_{d+1}^2$ fluctuating on the spacetime defined by \eqref{metrexample}. Let us write the Laplace operator in Fefferman-Graham and AdS-slicing focusing on $\rho$ and $u$-dependence correspondingly:
%\bea
%\Box \phi &= \frac{1}{L^2} \rho^2 \pa_{\rho}^2 \phi + \frac{(1-d)}{L^2} \rho \pa_{\rho} \phi + \ldots  = M_{d+1}^2 \phi \\
%& = \frac{1}{ \g^2 L^2 \cosh^2(z/L)} u^2 \pa_{u}^2 \phi + \frac{(2-d)}{ \g^2 L^2 \cosh^2(z/L)} u \pa_{u} \phi + \ldots
%\eea
%The functions $f(z)$ and $g(z)$ satisfy 
%\bea
%g'(z)^2 = \frac{1}{(d-1)^2 q(z) \Big(\frac{(d-1)^2}{L^2}q(z) -1 \Big)}, \\
%f'(z) = \Big( 1 - \frac{(d-1)^2}{L^2}q(z) \Big) g'(z).
%\eea

%%%%%%%%%%%%%%%%%%%%%%%%%%%%%%%%%%%%%%%%%%%%%%%%%%%%%%%%%%%%%
%% BIBLIOGRAPHY AND OTHER LISTS
%%%%%%%%%%%%%%%%%%%%%%%%%%%%%%%%%%%%%%%%%%%%%%%%%%%%%%%%%%%%%
%% A small distance to the other stuff in the table of contents (toc)
\addtocontents{toc}{\protect\vspace*{\baselineskip}}

%% The Bibliography
%% ==> You need a file 'literature.bib' for this.
%% ==> You need to run BibTeX for this (Project | Properties... | Uses BibTeX)
%\addcontentsline{toc}{chapter}{Bibliography} %'Bibliography' into toc
%\nocite{*} %Even non-cited BibTeX-Entries will be shown.
%\bibliographystyle{unsrt} %Style of Bibliography: plain / apalike / amsalpha / ...
%\bibliography{thesis} %You need a file 'thesis.bib' for this.
%% <== End of hints
%%%%%%%%%%%%%%%%%%%%%%%%%%%%%%%%%%%%%%%%%%%%%%%%%%%%%%%%%%%%%

%%%%%%%%%%%%%%%%%%%%%%%%%%%%%%%%%%%%%%%%%%%%%%%%%%%%%%%%%%%%%
%% BIBLIOGRAPHY AND OTHER LISTS
%%%%%%%%%%%%%%%%%%%%%%%%%%%%%%%%%%%%%%%%%%%%%%%%%%%%%%%%%%%%%

%%%%%%%%%%%%%%%%%%%%%%%%%%%%%%%%%%%%%%%%%%%%%%%%%%%%%%%%%%%%%
%% APPENDICES
%%%%%%%%%%%%%%%%%%%%%%%%%%%%%%%%%%%%%%%%%%%%%%%%%%%%%%%%%%%%%

%% ==> Write your text here or include other files.

%\input{FileName} %You need a file 'FileName.tex' for this.
%% A small distance to the other stuff in the table of contents (toc)
\addtocontents{toc}{\protect\vspace*{\baselineskip}}

%% The Bibliography
%% ==> You need a file 'literature.bib' for this.
%% ==> You need to run BibTeX for this (Project | Properties... | Uses BibTeX)
%\addcontentsline{toc}{chapter}{Bibliography} %'Bibliography' into toc
%\nocite{*} %Even non-cited BibTeX-Entries will be shown.
%\bibliographystyle{jhep} %Style of Bibliography: plain / apalike / amsalpha / ...
%\bibliography{literature} %You need a file 'literature.bib' for this.

\providecommand{\href}[2]{#2}\begingroup\raggedright\endgroup

%% The List of Figures

\end{document}